\documentclass[12pt]{article}

\usepackage{amssymb,amsfonts,amsthm}
\usepackage{amsmath}
\usepackage{bm}             
\usepackage[mathscr]{eucal}
\usepackage{cancel}
\usepackage{wasysym}
\usepackage{array}

\usepackage{graphicx}
\usepackage{subfigure}

\def\be{\begin{equation}}
\def\ee{\end{equation}}
\def\bea{\begin{eqnarray}}
\def\eea{\end{eqnarray}}

\def\hsp5{\hspace{5mm}}

\DeclareMathOperator{\sech}{sech}

\theoremstyle{remark}

\title{\sc Quintessence from a state space perspective}

\begin{document}

\author{
\sc Artur Alho,$^{1}$\thanks{Electronic address:{\tt
aalho@math.ist.utl.pt}}\,, Claes Uggla,$^{2}$\thanks{Electronic address:{\tt
claes.uggla@kau.se}}\,  and John Wainwright$^{3}$\thanks{Electronic
address:{\tt jwainwri@uwaterloo.ca}}\\
$^{1}${\small\em Center for Mathematical Analysis, Geometry and
Dynamical Systems,}\\
{\small\em Instituto Superior T\'ecnico, Universidade de Lisboa,}\\
{\small\em Av. Rovisco Pais, 1049-001 Lisboa, Portugal.}\\
$^{2}${\small\em Department of Physics, Karlstad University,}\\
{\small\em S-65188 Karlstad, Sweden.}\\
$^{3}${\small\em Department of Applied Mathematics, University of Waterloo,}\\
{\small\em Waterloo, ON, N2L 3G1, Canada.}}


\date{}
\maketitle

\begin{abstract}


We use dynamical systems methods to study quintessence models 
in a spatially flat and isotropic spacetime with matter and a scalar field 
with potentials for which $\lambda(\varphi)=-V_{,\varphi}/V$ is bounded, 
thereby going beyond the exponential potential for which 
$\lambda(\varphi)$ is constant. The scalar field equation of state parameter 
$w_\varphi$ plays a central role when comparing quintessence models with
observations, but with the dynamical systems used to date 
$w_\varphi$ is an indeterminate, discontinuous, function on the 
state space in the asymptotically matter dominated regime. 
Our first main result is the introduction of new variables that lead to a 
\emph{regular} dynamical system on a \emph{bounded} three-dimensional state 
space on which $w_\varphi$ is a \emph{regular} function. The solution trajectories 
in the state space then provide a visualization of different types of quintessence
evolution, and how initial conditions affect the transition between the matter 
and scalar field dominated epochs; this is complemented by graphs 
$w_\varphi(N)$, where $N$ is the $e$-fold time, which enables
characterizing different types of quintessence evolution.


\end{abstract}

\newpage

\section{Introduction\label{sec:intro}}

In 1998 observations of type Ia supernovae indicated that the
Universe is undergoing late-time acceleration~\cite{rieetal98,peretal99}.
Within the framework of General Relativity this cosmic acceleration implies that
there exists an exotic energy component in the Universe, called dark energy,
with an equation of state satisfying $w_\mathrm{DE}< -1/3$. Observational
constraints on $w_\mathrm{DE}$, assumed to be constant, provided by the cosmic
microwave background and the large-scale structure of the Universe continue to
tighten and are typically of order $-1.1<w_\mathrm{DE}<-0.9$ (see, e.g.,
Suzuki \emph{et al}. (2012)~\cite{suzetal12}, section 5,
Ade \emph{et al}. (2016)~\cite{adeetal16}, section 5, and more briefly,
Chiba \emph{et al}. (2013)~\cite{chietal13}, the introduction and section III).
The simplest hypothesis compatible with these observations is that the dark
energy is a cosmological constant $\Lambda$ with $w_\mathrm{DE}$ exactly
equal to $-1$, and that the Universe can be described by the $\Lambda$CDM model,
which is currently viewed as the standard model of cosmology. Although this simple
model is in surprisingly good agreement with observations, it is still possible
that the equation of state of dark energy is not constant but instead
depends on time in a way that is compatible with observations. For this reason other
candidates for dark energy have been extensively investigated. The simplest of
these is a dynamical canonical scalar field $\varphi$, minimally coupled to
gravity and with a potential $V(\varphi)$, referred to as
quintessence, the fifth element of the current matter content in the
Universe, after baryons, dark matter, radiation, and neutrinos,
Caldwell \emph{et al}. (1998)~\cite{caletal98}.

The quintessence mechanism generating the present acceleration of the
Universe resembles that of inflation in the very early Universe,
{\it i.e.}, a canonical scalar field rolling slowly down a slowly varying
potential $V(\varphi)$, although the energy scale of the quintessence
potential is tiny compared to that of the inflaton potential. Furthermore,
in contrast to inflation, quintessence dynamics also involve non-relativistic
matter (baryons and cold dark matter). Loosely speaking, the dynamics of
the quintessence field is governed by its equation of state parameter
$w_\mathrm{DE} = w_{\varphi}=p_{\varphi}/\rho_{\varphi}$, with
observations requiring that at the present time $w_{\varphi}$ is
sufficiently close to $-1$. We note that the $\Lambda$CDM model can be
viewed as a limiting quintessence model with constant potential $V(\varphi)=\Lambda$
and a constant scalar field, which results in $w_{\varphi}=-1$.
It is expected that upcoming large scale structure surveys will impose
increasingly tight constraints on the value of $w_{\varphi}$ and its time
rate of change at the present time (Akrami \emph{et al}. (2020)~\cite{akretal20}, see
the abstract).

In 1998 it was shown by Copeland \emph{et al.}~\cite{copetal98} that for a model
with a single matter component with a linear equation of state, and
with the simplest type of potential, the exponential potential, the
governing equations could be formulated as a dynamical system on a two dimensional
state space.\footnote{A dynamical system is a system of differential equations
of the form ${\bf x}'={\bf f}(\bf x),$ where ${\bf x}\in {\mathbb R}^n$ describes
the state space, ${\bf f}(\bf x)$ is a vector field on ${\mathbb R}^n,$
and ${}^\prime$ denotes the derivative with respect to time. Dynamical systems methods
were first used in cosmology in 1971 by Collins~\cite{collins71}, using two-dimensional
systems to study anisotropic Bianchi models. This early work has subsequently
been extended by many researchers to systems in higher dimensions, see
e.g.~\cite{waiell97,col03,bahetal18} and references therein.}
This enabled cosmologists to apply standard techniques of dynamical systems
theory to describe the evolution of these simple quintessence models.
For more general potentials, however, one needs a three-dimensional autonomous
system of equations in order to describe the extra degree of freedom in the scalar
field.

In this paper we consider potentials such that $\lambda(\varphi)=-V_{,\varphi}/V$
is bounded, the simplest being the exponential potential for which $\lambda$ is constant.
To study models with varying $\lambda(\varphi)$, Alho and Uggla (2015)~\cite{alhugg15b} introduced a
regular dynamical system on a bounded three-dimensional state space, where two of the
variables were closely related to those used by Copeland \emph{et al.}~\cite{copetal98}
for an exponential potential. As will be shown in a future paper, this system
is effective for deriving new simple and accurate approximations of scalar
field quantities such as the equation of state parameter $w_\varphi$ and the
Hubble-normalized scalar field energy density $\Omega_\varphi=\rho_\varphi/3H^2$,
thereby simplifying comparisons with observational data. However, it has the
drawback that $w_\varphi$ is indeterminate in the asymptotically matter
dominated regime where $\Omega_\varphi$ is zero. We therefore introduce
two new variables, closely related to $w_\varphi$ and $\Omega_\varphi$,
that lead to a \emph{new regular dynamical system on a bounded three-dimensional
state space} that \emph{avoids the indeterminacy problem}, which allows us
to define and study various types of quintessence.

According to Tsujikawa (2013)~\cite{tsu13} there are three types of
quintessence in the literature: thawing, scaling freezing,
and tracking freezing quintessence. The concepts of thawing
and freezing were defined by Caldwell and Linder (2005)~\cite{callin05} as
follows: thawing is characterized by $w_\varphi\approx - 1$
where $w_\varphi$ subsequently grows, {\it i.e.} $w_\varphi^\prime>0$, while
$w_\varphi>-1$ and $w_\varphi^\prime<0$ holds for freezing.\footnote{Here 
$^\prime$ denotes the derivative with respect to $e$-fold time $N$,
see equation~\eqref{Ndef}.} In this paper
we will refine the classification in~\cite{tsu13} and
in addition to thawing, scaling freezing, and tracking freezing quintessence
also define freezing quintessence, scaling oscillatory
quintessence and oscillatory quintessence.

The outline of the paper is as follows. In the next section we derive the new
dynamical system, assuming a potential with bounded $\lambda(\varphi)$. In
section~\ref{fut.att.conj} we describe global properties of the
state space and discuss invariant boundary sets, fixed points and asymptotic behaviour.
Section~\ref{sec:overview} contains a discussion of various types of quintessence
from a state space perspective using the new dynamical system. The following section
uses the double-exponential potential as a simple example to illustrate the different
types of quintessence by means of state space pictures and complementary graphs of
$w_\varphi(N)$ and $H_\varphi(N)/H_\Lambda(N)$, {\it i.e.} the ratio of the Hubble
variable for quintessence and $\Lambda$CDM. Section~\ref{concl.remarks}
summarizes our
results, illustrating our quintessence classification scheme
with examples from the literature, and concludes with some comments about future
developments.

\section{Derivation of the new dynamical system\label{sec:derivationdyn}}

Consider a spatially flat and isotropic Friedmann-Lema\^{i}tre-Robertson-Walker
(FLRW) spacetime,
\begin{equation}
ds^2 = -dt^2 + a^2(t)\delta_{ij}dx^idx^j,
\end{equation}
where $a(t)$ is the cosmological scale factor. The source consists of
matter with an energy density $\rho_\mathrm{m}>0$ and pressure
$p_\mathrm{m}=0$, which represents cold dark matter\footnote{This 
simple model is useful for describing the transition from
an epoch of matter domination to an epoch in which the scalar field is dominant.
A more realistic model is provided by a two component source with
matter and radiation, in which there is an epoch of radiation domination
preceding the epoch of matter domination, see the end of section~\ref{LCDMsec}.
Although generalizing the discussion to include radiation is straightforward, 
it leads to a four-dimensional state space instead of the present three-dimensional 
one, which complicates visualization. For simplicity and pedagogically illustrative 
reasons we therefore neglect radiation in this paper, although we will 
include it in future work.}
and a minimally coupled scalar field, $\varphi$, with a
potential $V(\varphi)>0$, which results in
\begin{equation}\label{rhophipphi}
\rho_\varphi = \frac12\dot{\varphi}^2 + V(\varphi),\qquad
p_\varphi = \frac12\dot{\varphi}^2 - V(\varphi),
\end{equation}
where an overdot represents the derivative with respect to the cosmic
proper time $t$. The Raychaudhuri equation, the Friedmann equation, the (non-linear) Klein-Gordon
equation, and the energy conservation law for matter with zero pressure,
can be written as\footnote{We use units so
that $c=1$ and $8\pi G=1$, where $c$ is the speed of light and
$G$ is the gravitational constant.}
\begin{subequations}\label{Mainsysdim}
\begin{align}
\dot{H} + H^2 &= -\frac16(\rho + 3p), \label{Ray}\\
3H^2 &= \rho, \label{Gauss}\\
\ddot{\varphi} &=-3H\dot{\varphi} - V_{,\varphi}, \label{KG}\\
\dot{\rho}_\mathrm{m} &= -3H\rho_\mathrm{m},
\end{align}
\end{subequations}
where the Hubble variable is defined by $H=\dot a/a$, and the total energy density
$\rho$ and pressure $p$ are given by
\begin{equation}
\rho = \rho_\varphi + \rho_\mathrm{m},\qquad
p = p_\varphi.
\end{equation}
Since $\rho>0$ implies that $H^2>0$ in~\eqref{Gauss}, it follows that $H>0$ for
initially expanding models.

%
%

To obtain useful dynamical systems, we first introduce the following
\emph{dimensionless} and \emph{bounded} Hubble-normalized
quantities:\footnote{The variable
$\Sigma_\varphi$ was first introduced by Coley
\emph{et al.} (1997)~\cite{coletal97} and Copeland
\emph{et al.} (1998)~\cite{copetal98}, whose $x$ is $\Sigma_\varphi$.
Since then, $\Sigma_\varphi$ (or $\varphi^\prime$)
is now commonly used to describe scalar fields in cosmology,
see, e.g., Urena-Lopez (2012)~\cite{ure12}, equation (2.3),
Tsujikawa (2013)~\cite{tsu13}, equation (16) and Alho and Uggla
(2015)~\cite{alhugg15b}, equation (8). The reason for using the
notation $\Sigma$ for the kernel is because $\Sigma_\varphi$ plays
a role that is similar to Hubble-normalized shear, which is typically
denoted with the kernel $\Sigma$, see e.g.~\cite{waiell97}.}
\begin{subequations}\label{vardef1}
\begin{align}
\Sigma_\varphi &\equiv \frac{\dot{\varphi}}{\sqrt{6}H} = \frac{\varphi^\prime}{\sqrt{6}}, \label{Sigvarphidef}\\
\Omega_V &\equiv \frac{V}{3H^2},\label{OmVdef}\\
\Omega_\mathrm{m} &\equiv \frac{\rho_\mathrm{m}}{3H^2}.\label{Ommdef}
\end{align}
\end{subequations}
A ${}^\prime$ henceforth denotes the derivative with respect to $e$-fold time
\begin{equation}\label{Ndef}
N \equiv \ln(a/a_0),
\end{equation}
where $t=t_0 \Rightarrow N=0$ refers to the present time, and $a_0 = a(t_0)$.
The definition~\eqref{Ndef} implies that $N\rightarrow - \infty$
and $N\rightarrow + \infty$ when $a\rightarrow 0$ and $a \rightarrow\infty$, respectively.

To obtain dimensionless dynamical systems we replace the cosmic proper time $t$ with
the dimensionless $e$-fold time $N$ using the following relations:
\begin{equation}
\frac{d}{dt}=H\frac{d}{dN},\qquad \frac{d^2}{dt^2}=
H^2\left(\frac{d^2}{dN^2} - (1+q)\frac{d}{dN}\right),
\end{equation}
where
\begin{equation}\label{qdef1}
q \equiv -\frac{a\ddot{a}}{\dot{a}^2} = -1 - \frac{H^\prime}{H}
\end{equation}
is the \emph{deceleration parameter}.

We consider potentials $V(\varphi)$ for which
\begin{subequations} \label{lambda.pm}
\begin{equation} \label{lambda.def}
\lambda(\varphi) \equiv -\frac{V_{,\varphi}}{V}
\end{equation}
is a regular bounded function for all $\varphi$ with limits
\begin{equation} \label{lambda.lim}
\lim_{\varphi\rightarrow\pm\infty}\lambda = \lambda_\pm,
\end{equation}
\end{subequations}
\emph{i.e.}, we consider asymptotically exponential (or constant, when $\lambda_\pm=0$) potentials,
since $V \propto \exp(-\lambda\varphi)\, \Rightarrow\, \lambda(\varphi) = \lambda$.
Depending on the form of the potential we choose a regular bounded increasing
(and hence invertible) function $\bar{\varphi}(\varphi)$ which satisfies
\begin{equation} \label{lim.varphi}
\lim_{\varphi\rightarrow\pm\infty} \bar{\varphi}(\varphi) = \pm 1, \qquad
\lim_{\varphi\rightarrow\pm\infty}d\bar{\varphi}/d\varphi=0.
\end{equation}
For a fairly wide class of simple potentials a suitable choice of
$\bar\varphi$ is\footnote{More generally the choice of $\bar{\varphi}$
should be adapted to the properties of $\lambda(\varphi)$, especially
the asymptotic ones; for some examples,
see Alho and Uggla (2015)~\cite{alhugg15b}.}
\begin{equation} \label{def.bar.phi}
\bar\varphi (\varphi) = \tanh(C\varphi + D) \quad\implies \quad
\frac{d\bar\varphi}{d\varphi} = C(1-\bar\varphi^2),
\end{equation}
where the choice of the constants $C>0$ and $D$ depends on the potential.
Since $\lambda(\varphi)$ is assumed to be a bounded and sufficiently
differentiable function of $\varphi \in {\mathbb R}$ and the function
$\bar\varphi(\varphi)$ is invertible we can define (with a slight abuse of notation)
$\lambda(\bar\varphi)=\lambda(\varphi(\bar\varphi))$, where $\varphi(\bar\varphi)$
is the inverse function of $\bar\varphi (\varphi)$. It follows from~\eqref{lambda.lim}
and~\eqref{lim.varphi} that
\begin{equation}
\lambda(\bar{\varphi} = \pm 1) = \lambda_\pm.
\end{equation}

Using $({\bar\varphi}, \Sigma_{\varphi},\Omega_\mathrm{m})$ as the
state vector and $N$ as the time variable, the definitions~\eqref{vardef1},~\eqref{qdef1},
and the equations in~\eqref{Mainsysdim} result in the following dynamical
system:\footnote{This system was used by Alho
and Uggla (2015)~\cite{alhugg15b}, equation (9), with variables
$(x,\Omega_{\mathrm m}, Z)$, where $x\equiv\Sigma_{\varphi}$ while $Z$
corresponds to ${\bar\varphi}$.}
\begin{subequations}\label{Dynsys1}
\begin{align}
\bar{\varphi}^\prime &= \sqrt{6}\left(\frac{d\bar{\varphi}}{d\varphi}\right)\Sigma_\varphi
=\sqrt{6}C(1-\bar{\varphi}^2)\Sigma_\varphi,\label{varphi}\\
\Sigma_\varphi^\prime &= - (2-q)\Sigma_\varphi +
\sqrt{\frac32}\lambda(\bar{\varphi})\Omega_V, \label{Sigmaeqmatter1} \\
\Omega^{\prime}_\mathrm{m} &= (2q - 1)\Omega_\mathrm{m}, \label{Ommeq1}
\end{align}
where
\begin{align}
q &= -1 + 3\Sigma^2_\varphi + \frac32\Omega_\mathrm{m},\label{qdef}\\
\Omega_V &= 1 - \Sigma_\varphi^2 - \Omega_\mathrm{m}.\label{OmVeq}
\end{align}
\end{subequations}
The state space is bounded and described by the
inequalities:
\begin{equation}\label{bounds.ss}
-1 \leq \bar\varphi\leq 1,\qquad \Omega_\mathrm{m} \geq 0, \qquad
\Omega_V = 1 - \Sigma_\varphi^2 - \Omega_\mathrm{m} \geq 0,
\end{equation}
which follow from $\rho_\mathrm{m}\geq 0$ and $V\geq 0$.
Specifying the scalar field potential and thereby
$\lambda(\bar\varphi)$ results in that~\eqref{Dynsys1} forms a
\emph{regular dynamical system on a bounded state space}
for a wide class of scalar field
models.\footnote{That the dynamical system
${\bf x}'={\bf f}(\bf x)$ is regular means that
the functions in the vector field on the right side are
bounded and sufficiently differentiable on the state space.
If a dynamical system is regular one can use standard techniques
to investigate the stability of the fixed points
(singular points, critical points, equilibrium points).
A bounded state space has the advantage that
a regular dynamical system has future and past
attractors in the state space and that one can
describe the global properties of the solutions.}
Due to~\eqref{Dynsys1}, $\Omega_V$ obeys the auxiliary equation
\begin{equation}
\Omega_V^\prime = 2\left(1 + q - \sqrt{\frac32}\lambda(\bar{\varphi})\Sigma_\varphi\right)\Omega_V,
\end{equation}
from which it follows that $\Omega_V=0$ is an invariant boundary set,
as is $\Omega_\mathrm{m}=0$ due to~\eqref{Ommeq1}.

\begin{figure}[ht!]
	\begin{center}
     \subfigure[The `tent state space' .
      ]{\label{TentSS}
			\includegraphics[width=0.30\textwidth]{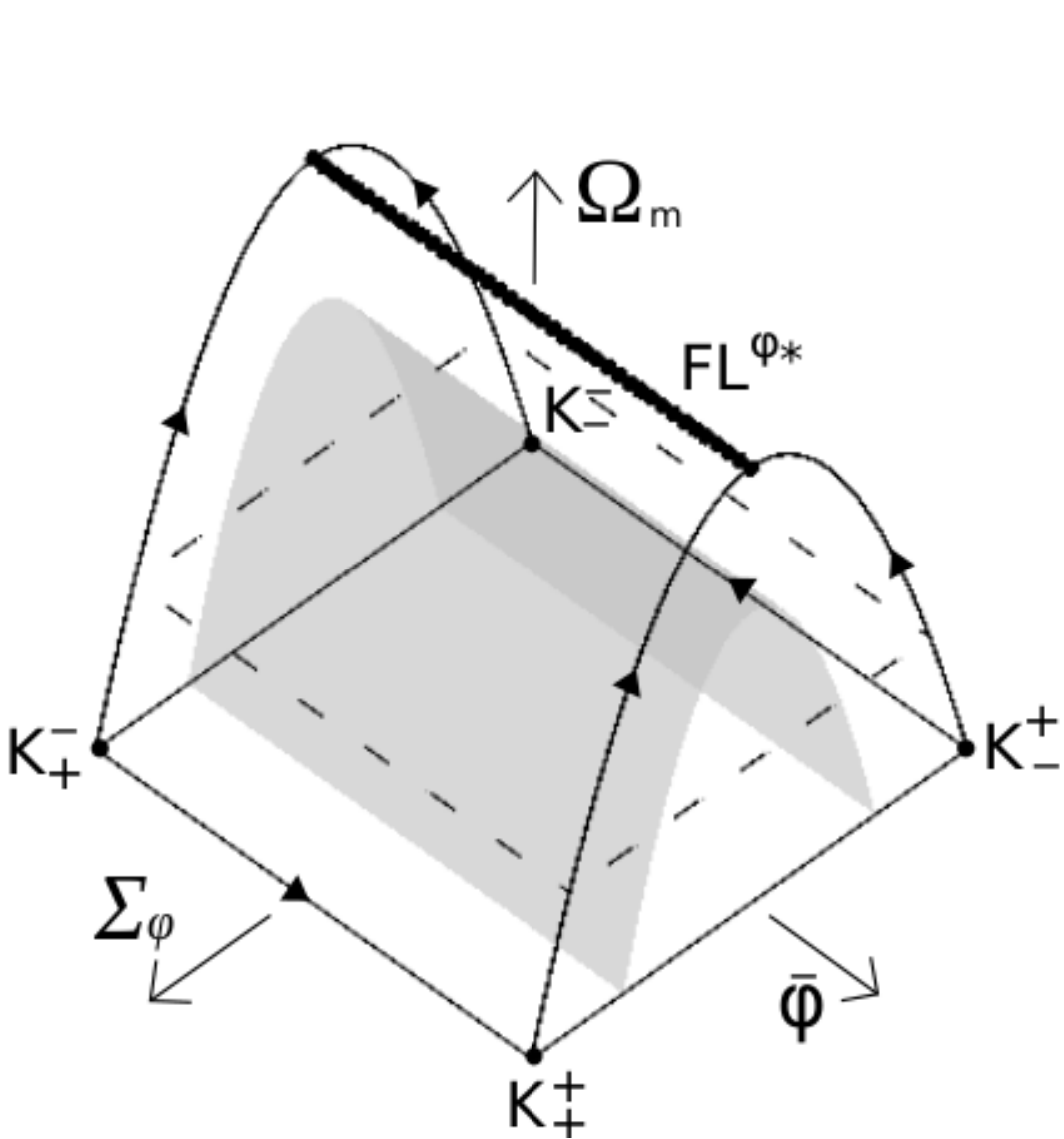}}
		\subfigure[The `box state space'.
		]{\label{BoxSS1}
			\includegraphics[width=0.30\textwidth]{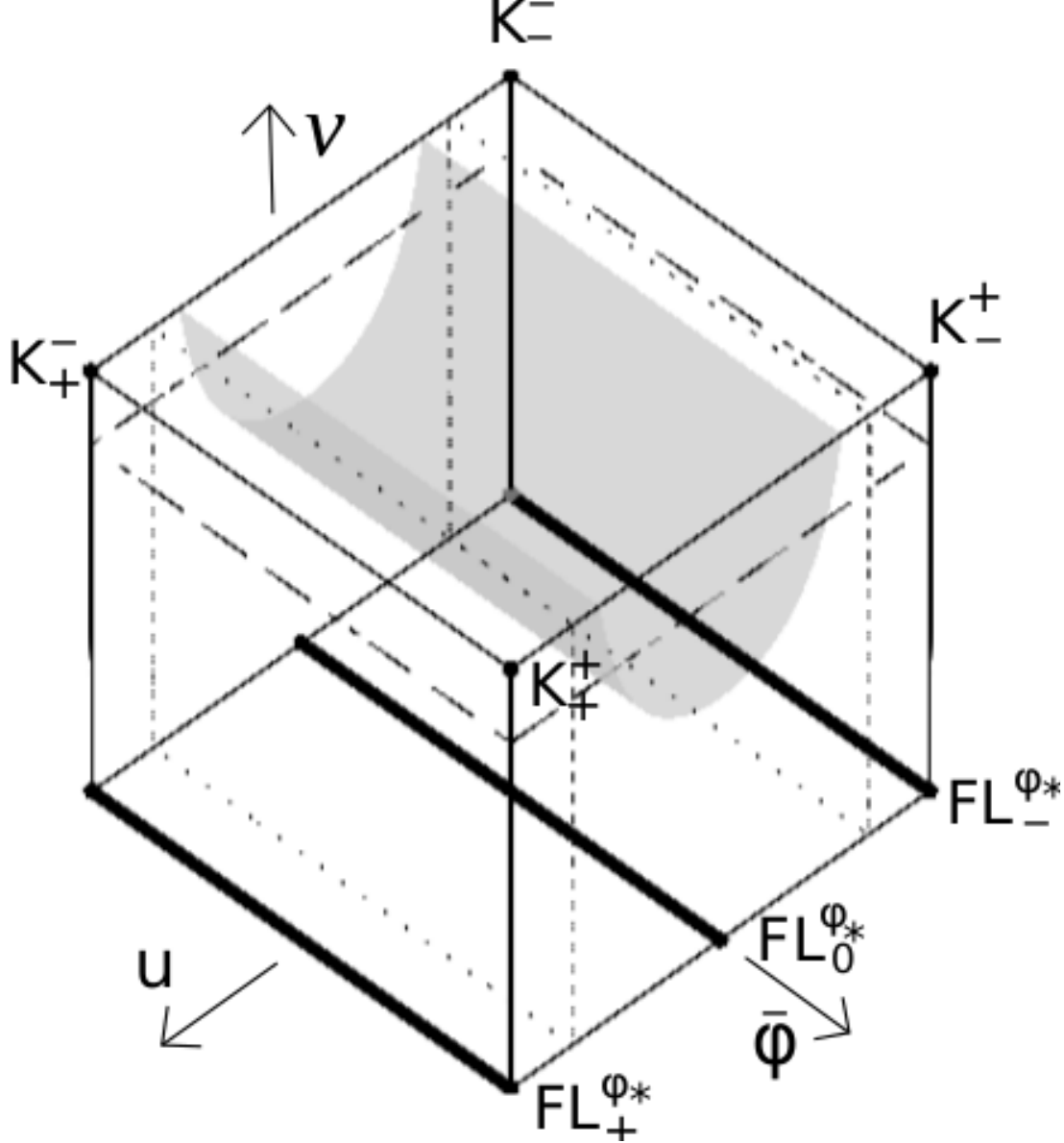}}
		\subfigure[Model independent orbits.
		]{\label{BoxSS2}
			\includegraphics[width=0.30\textwidth]{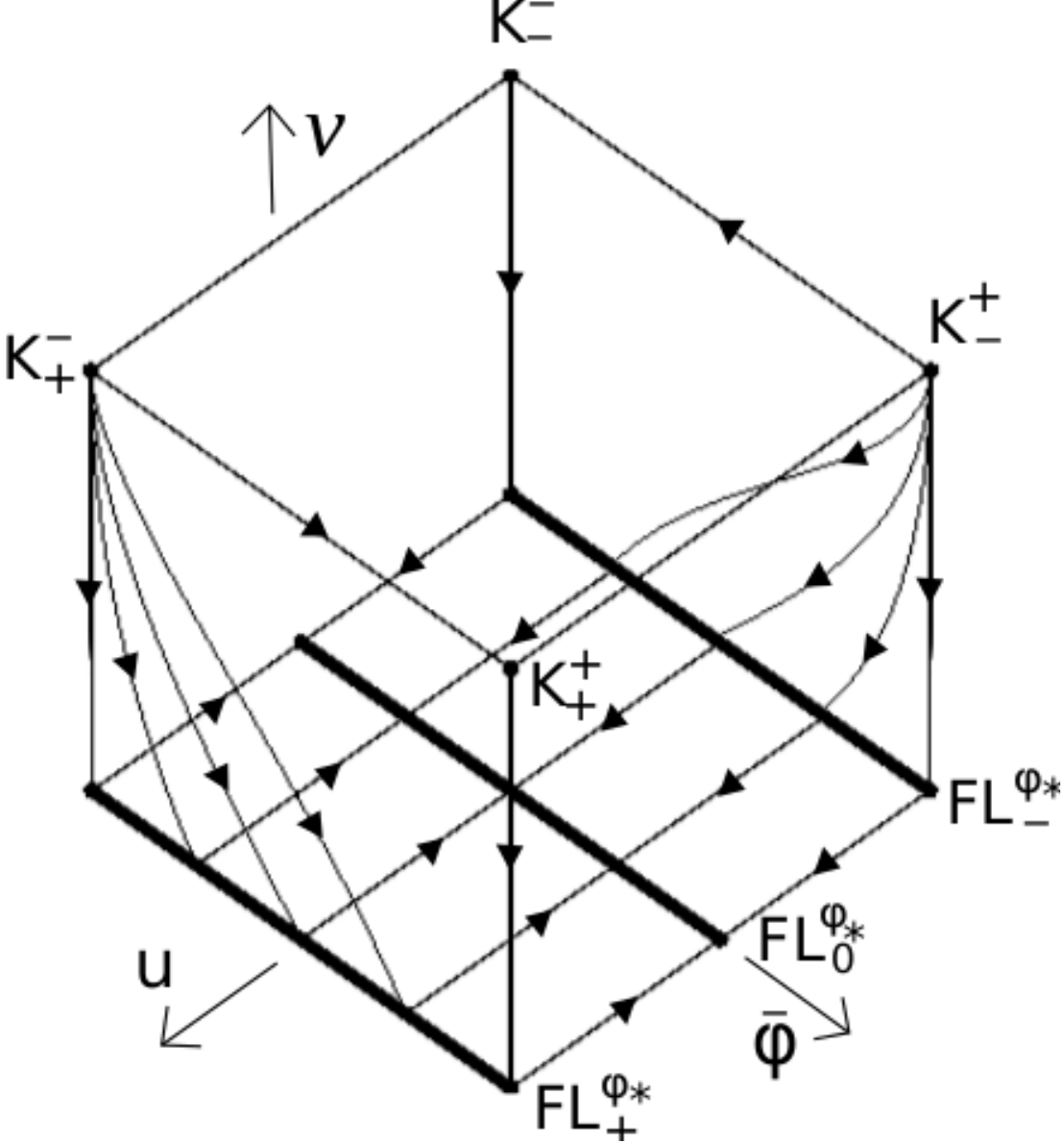}}
			\vspace{-0.5cm}
	\end{center}
\caption{The `tent' and `box state spaces'. The horizontal dashed lines in (a) and (b) depict when $\Omega_\varphi=0.68$,
located at $v\approx 0.48 = 0.82v_\mathrm{max}$, $v_\mathrm{max} = 1/\sqrt{3}$, while the gray parabolic
surfaces correspond to $q=0$. The dotted lines in (b) depict the planes $u=\pm1$ on which $w_\varphi=0$
and $v^\prime = 0$. Figure (c) shows the orbit structure on the model independent boundaries
$u=\pm\sqrt{2}$ ($w_\varphi = 1$; $\Omega_V=0$) and $v=0$ ($\Omega_\varphi=0$, $\Omega_\mathrm{m}=1$).}
\label{SSTentBox}
\end{figure}

The inequalities in~\eqref{bounds.ss} imply that
$\Omega_\mathrm{m}$ and $\Sigma_\varphi^2$ are bounded above by $1$.
The state space can be visualized as a bounded three-dimensional
set with rectangular base
$-1\leq\Sigma_\varphi\leq 1$, $\Omega_\mathrm{m}=0$, $-1\leq\bar{\varphi} \leq 1$,
bounded above by the tent-like surface $\Sigma_\varphi^2 + \Omega_\mathrm{m}=1$,
and with vertical ends $\bar\varphi=\pm1$, see Figure~\ref{TentSS}.

Before continuing we digress to place the dynamical
system~\eqref{Dynsys1} in a historical context. If $\lambda$ is
constant, which means that the potential is exponential,
$V \propto \exp(-\lambda \varphi)$, equations \eqref{Sigmaeqmatter1}
and~\eqref{Ommeq1} form a two-dimensional dynamical system
that is closely related to the widely used $(x,y)$ system introduced by
Copeland \emph{et al}. (1998)~\cite{copetal98} in which the dynamical variables
are\footnote{See their equation (5) and our equation~\eqref{vardef1}.
To obtain~\eqref{Dynsys1} from their equations, use $y^2=1-\Sigma_\varphi^2-\Omega_\mathrm{m}$
and $\Omega_\mathrm{m}$ as the dynamical variable.}
$x=\Sigma_\varphi$ and $y=\sqrt{\Omega_V}$.
To go beyond the exponential potential, $\lambda$ is sometimes added as a dynamical variable,
whose evolution equation is given by
\begin{equation}
\lambda'=-\sqrt6 \Sigma_\varphi \lambda^2(\Gamma(\varphi)-1),\qquad
\Gamma(\varphi) = \frac{V V_{,\varphi\varphi}}{V_{,\varphi}^2}.
\end{equation}
The resulting system is not autonomous due to the dependence of $\lambda$
and $\Gamma$
on $\varphi$. However, if $\lambda(\varphi)$ is monotone and hence invertible,
\emph{i.e.}  we can write $\varphi=\varphi(\lambda)$ it follows that $\Gamma$
is a function of $\lambda$:
$\Gamma=\Gamma(\varphi(\lambda))$. For this restricted class of potentials the
resulting system using $\lambda$ as a variable is autonomous.\footnote{See 
Bahamonde \emph{et al}. (2018)~\cite{bahetal18},
page 40, equations (4.53)-(4.56), and references given there. In
particular, Table 10 in~\cite{bahetal18} gives a  list of potentials for
which $\Gamma$ can be written as a function of $\lambda$, but for most of
these potentials $\lambda$ is unbounded. This system of equations has
been used, e.g., by Fang \emph{et al}. (2009)~\cite{fanetal09},
see their equations (5)-(7), and Urena-Lopez (2012)~\cite{ure12},
equations (2.4) and (2.8).}

In setting up the dynamical system~\eqref{Dynsys1} 
we have avoided imposing this restriction on the potential by initially
\emph{simply adding $\varphi$ as a variable}, following Nunes and Mimoso (2000)~\cite{nunmim00}
and Alho and Uggla (2015)~\cite{alhugg15b}, with the definition~\eqref{Sigvarphidef}
of $\Sigma_\varphi$ acting as the evolution equation for $\varphi$, and then replacing
$\varphi$ with a suitable bounded variable $\bar{\varphi}$.

We next introduce two key quantities associated with the
scalar field: the \emph{Hubble-normalized scalar field energy density}
$\Omega_\varphi$ and the \emph{equation of state parameter}
$w_\varphi$,
\begin{subequations} \label{var.scalar}
\begin{align}
\Omega_\varphi &\equiv \frac{\rho_\varphi}{3H^2} = \Sigma_\varphi^2 + \Omega_V = 1-\Omega_\mathrm{m}, \\
w_\varphi &\equiv \frac{p_\varphi}{\rho_\varphi} =  \frac{\Sigma_\varphi^2 - \Omega_V}{\Sigma_\varphi^2 + \Omega_V}
= -1+\frac{2\Sigma_\varphi^2}{\Omega_\varphi},\quad
\text{provided that}\quad \Omega_\varphi>0. \label{w.varphi}
\end{align}
\end{subequations}
It follows that $-1\leq w_\varphi \leq 1$ where
$w_\varphi =1$ when $\Omega_V=0$, $\Sigma_\varphi\neq 0$, and
$w_\varphi =-1$ when $\Sigma_\varphi=0$, $\Omega_\varphi>0$.
These quantities, while not directly observable, play an important role
in determining the restrictions observations place on quintessence evolution.
It follows from~\eqref{var.scalar}, however, that $w_\varphi$ \emph{is indeterminate
when} $\Omega_\varphi=0\,\Rightarrow\, \Sigma_\varphi=0$, which is a drawback of
using~\eqref{Dynsys1} as evolution equations. To avoid this problem one could
consider using $\Omega_\varphi$ and $w_\varphi$ as dynamical variables.
It follows from~\eqref{Dynsys1} that the evolution equations for
$\Omega_\varphi$ and $w_\varphi$ are given by
\begin{subequations}\label{alt.evol.eqs}
\begin{align}
w_\varphi' &= -3(1- w_\varphi)
\left(1+w_\varphi-\sqrt{\frac{2}{3}}\lambda(\bar{\varphi})\Sigma_\varphi\right),\\
\Omega_\varphi' &= -3w_\varphi(1-\Omega_\varphi)\Omega_\varphi.\label{alt.evol.eqs.2}
\end{align}
\end{subequations}
This is \emph{not} a regular system due to the appearance of
$\Sigma_\varphi$ which has to be obtained from~\eqref{w.varphi}.\footnote{Bahamonde
\emph{et al}. (2018)~\cite{bahetal18}, page 26, comment that although
$\Omega_\varphi$ and $w_\varphi$ are useful when comparing with observational
data, there are mathematical difficulties in using them as dynamical variables.}
This appears to be a dead-end as regards obtaining regular evolution equations
in a state space on which $w_\varphi$ is well-defined. However, we have noticed that
the system~\eqref{alt.evol.eqs} can be made regular by writing
\begin{equation} \label{reg.var}
\Omega_\varphi = 3v^2, \qquad w_\varphi = u^2 - 1,
\end{equation}
and specifying that $u$ has the same sign as $\Sigma_\varphi$ and that $v\geq 0$,
which with~\eqref{w.varphi} implies the following key relations:\footnote{The transformation~\eqref{tent.to.box.map}
maps the two-dimensional boundary set $v=0$ of the `box state space' onto the one-dimensional
line of fixed points $\Sigma_\varphi=0$, $\Omega_\varphi=0$, $\Omega_\mathrm{m}=1$ in
the `tent state space', and is thereby not one-to-one, see Figure~\ref{SSTentBox}. The
factor $\sqrt{2}$ in the definition of $u$ leads to the simple relation
$w_\varphi = u^2 -1$ while the factor $1/\sqrt{3}$ in the definition of $v$
avoids a factor $\sqrt{3}$ in the equations.}
\begin{subequations} \label{tent.to.box.map}
\begin{alignat}{2}
u &\equiv \Sigma_\varphi\sqrt{\frac{2}{\Omega_\varphi}}, &\qquad
v &\equiv \sqrt{\frac{\Omega_\varphi}{3}},\\
\Sigma_\varphi &= \sqrt{\frac32}\,uv,&\qquad \Omega_\varphi &=3v^2.
\end{alignat}
\end{subequations}
It follows that the system~\eqref{Dynsys1} assumes the form
\begin{subequations}\label{Dynsys.uv}
\begin{align}
{\bar\varphi}^\prime &= 3\left(\frac{d\bar{\varphi}}{d\varphi}\right)uv
= 3Cuv(1-\bar{\varphi}^2),\label{barvarphiprime}\\
u^\prime &= \frac{3}{2}(2-u^2)(v\lambda(\bar{\varphi})-u), \label{u.prime} \\
v^\prime &= \frac{3}{2}(1-u^2)(1-3v^2)v, \label{v.prime}
\end{align}
\end{subequations}
where $C>0$ is the constant in the definition~\eqref{def.bar.phi} of $\bar\varphi$.
The state space is bounded and is described by the following inequalities:
\begin{equation}
-1\leq\bar\varphi\leq1,\qquad -\sqrt2\leq u\leq\sqrt2, \qquad
0\leq  v\leq  1/\sqrt3.
\end{equation}
Once the scalar field potential has been specified and
$\lambda(\bar\varphi)$ is determined, \eqref{Dynsys.uv}
forms a \emph{regular dynamical system on a bounded state space} for a wide
class of scalar field potentials with bounded $\lambda(\bar{\varphi})$,
where the equation of state parameter $w_\varphi=u^2-1$ is a regular function
on the state space.

Thinking of $(\bar\varphi,u,v)$ as Cartesian coordinates, the state space can be
visualized as a rectangular box, with $\bar{\varphi} = \pm 1$ describing the
invariant exponential potential boundaries, the invariant base $v=0$ and top $v=1/\sqrt3$
boundaries, representing $\Omega_\varphi=0, 1$, respectively, while the invariant
boundary sides $u=\pm\sqrt2$ correspond to $w_\varphi=1$. Due to this invariant boundary
structure we will refer to this state space as the \emph{`box state space'},
see Figure~\ref{BoxSS1}. 
Note that it is easy to visualize $w_\varphi$ and $\Omega_\varphi$
since $w_\varphi$ is constant on the vertical planes $u=\mathrm{constant}$ while
$\Omega_\varphi$ is constant on the horizontal planes $v=\mathrm{constant}$, as
follows from~\eqref{reg.var}.

Since
\begin{subequations}\label{qOmV}
\begin{align}
\Omega_V &= \frac{3}{2}(2-u^2)v^2,\label{omega.v}\\
q &= \frac12[1+9(u^2-1)v^2],
\label{q.v}
\end{align}
\end{subequations}
it follows that $\Omega_V=0$ on the $u=\pm\sqrt{2}$ ($w_\varphi=1$) boundaries while the region of
the box state space in which the orbits ({\it i.e.} solution trajectories) describe
accelerating models $(q<0)$ is a trough parallel to the $\bar\varphi$ axis with
a parabola-like profile given by $9(1-u^2)v^2 = 1$, which is shaded in grey
in Figure~\ref{BoxSS1}.
%
%
%
%

The form of the orbits on the invariant boundaries $v=0$ and $u=\pm\sqrt2$
is independent of $\lambda(\varphi)$, and thereby also of the potential. First,
the base of the box state space, $v=0$, forms the Friedmann-Lema\^{i}tre
$\mathrm{FL}$ invariant boundary set with $\Omega_\varphi=0$, $\Omega_\mathrm{m}=1$.
Since $\bar\varphi' = 0$ and $u^\prime = -\frac{3}{2}(2-u^2)u$ on this set
it follows that there are three lines of fixed points:
\begin{subequations}\label{FL.fixed}
\begin{alignat}{2}
\mathrm{FL}_0^{\varphi_*}&\!: &\qquad (\bar\varphi,u,v) &= (\bar{\varphi}_*,0,0),\\
\mathrm{FL}_{\pm}^{\varphi_*}&\!: &\qquad (\bar\varphi,u,v) &= (\bar{\varphi}_*,\pm\sqrt2,0),
\end{alignat}
\end{subequations}
with the constant $\bar{\varphi}_*$ satisfying $-1\leq \bar{\varphi}_*\leq1$,
where $w_\varphi=-1$ for $\mathrm{FL}_0^{\varphi_*}$ and
$w_\varphi=1$ for $\mathrm{FL}_{\pm}^{\varphi_*}$.
The lines of fixed points are connected by
heteroclinic orbits\footnote{A heteroclinic orbit
is an orbit that joins two different fixed points.}
$\mathrm{FL}_\pm^{\varphi_*} \rightarrow\mathrm{FL}_0^{\varphi_*}$ that are straight lines
with $\bar{\varphi} = \bar{\varphi}_* = \mathrm{constant}$, and thus the evolution of the scalar field
is `frozen' on $v=0$.
Note that the line of fixed points $\mathrm{FL}_0^{\varphi_*}$ and the orbits that originate from
$\mathrm{FL}_0^{\varphi_*}$ into the state space exist for \emph{all} models with matter
and a scalar field, and play a central role in our description of quintessence evolution.

Second, the orbits on the $u=\pm\sqrt2$ boundaries can also be determined explicitly,
as shown in~\cite{alhugg22,alhetal22}. In particular, all the orbits on the $u=\sqrt{2}$
($u=-\sqrt{2}$) boundary originate from the fixed point $\mathrm{K}^-_+$ ($\mathrm{K}_-^+$)
and end at the line $\mathrm{FL}_+^{\varphi_*}$ ($\mathrm{FL}_-^{\varphi_*}$),
where each fixed point attracts a single orbit, constituting its
stable manifold (properties of fixed points and motivation for their notation are
discussed in the next section). It is also helpful to note that
$\bar\varphi\in (-1,1)$ is increasing if $u>0$ and decreasing if $u<0$, which
{\it e.g.} determines the flow directions along the boundaries $u = \pm\sqrt{2}$.

The above features of the box state space, which are independent of the specific form of
the potential, are depicted in Figure~\ref{BoxSS2}.
%

In the next section we will derive global properties of orbits in the box
state space $(\bar{\varphi},u,v)$ using the dynamical system~\eqref{Dynsys.uv}.
This is accomplished by deriving a monotonic function and by describing the orbit structures
on the remaining model dependent boundaries. This then leads to a classification
of models with bounded $\lambda(\varphi)$ and
$\lim_{{\varphi}\rightarrow \pm\infty} = \lambda_\pm$ in terms of
their possible asymptotic features.

\section{Dynamical systems features and asymptotics\label{fut.att.conj}}

The interior box state space $(\bar{\varphi},u,v)$, {\it i.e.} the box state
space without its boundaries, can be viewed as being divided
into a central slab $-1<u<1$, in which $w_\varphi<0$ and where $\Omega_\varphi$ is
increasing,
and two outer slabs $1<u^2<2 $ in which $w_\varphi$ satisfies $0<w_\varphi<1$ and
where $\Omega_\varphi$ is decreasing, as follows from~\eqref{v.prime}.
In addition to these global features, the global dynamics of the
present models is severely restricted by the following monotonic function:
\begin{subequations}\label{3H2mon}
\begin{align}
3H^2 &= \frac{V(\bar{\varphi})}{\Omega_V} = \frac{2V(\bar{\varphi})}{3(2-u^2)v^2},\\
(3H^2)^\prime &= - 2(1+q)(3H^2) = -3(1 - 3v^2 + 3(uv)^2)(3H^2),
\end{align}
\end{subequations}
where we have used~\eqref{qdef1} and~\eqref{qOmV}.
Since $3H^2 = 2V(\bar{\varphi})/3(2-u^2)v^2$ is strictly monotonically
decreasing in the interior state space $(\bar{\varphi},u,v)$ due to that
$1 + q$ and $1 - 3v^2 + 3(uv)^2$ then are positive, it follows that there are
no interior fixed points or periodic orbits and that the asymptotics for all
interior orbits reside on the invariant boundary sets.

The different model dependent possibilities for the asymptotics of the interior orbits
for the present class of models are therefore determined by the $\bar{\varphi}=\pm 1$ and
$v=1/\sqrt{3}$ boundary sets, since it is only on these boundary sets where
$\lambda(\bar{\varphi})$ occurs in the boundary equations. Let us therefore begin
by considering these model dependent boundaries, which then naturally leads
to (A) a classification of the present models, and (B) the asymptotic features
for the different models this classification gives rise to.


\subsection{The scalar field dominant boundary set $v=1/\sqrt{3}\,\Leftrightarrow\, \Omega_\varphi=1$}

The intersection of the $v=1/\sqrt{3}$ ($\Omega_\varphi=1$; $\Omega_\mathrm{m} = 0$) and $u=\pm\sqrt{2}$
($w_\varphi = 1$; $\Omega_V=0$) boundaries contains four
\emph{kinaton fixed points}\footnote{The term kination, {\it i.e.}, kinetic scalar field
energy domination, was first introduced in~\cite{joypro98}
in the context of inflation and thereby in the absence of a matter field and hence
$\Omega_\varphi = \Sigma_\varphi^2 + \Omega_V = 1$. Kination then
corresponds to $\Sigma_\varphi = \pm 1$, $\Omega_V=0$, and thereby
$u=\pm\sqrt2$, $v=1/\sqrt{3}$, which results in $q=2$ and
$H^\prime = -(1+q)H= -3H \Rightarrow 3H^2 = \rho =\rho_\varphi = 3H_0^2\exp(-6N)
= 3H_0^2(a_0/a)^6$.} 
$\mathrm{K}^+_{\pm}$ and $\mathrm{K}^-_{\pm}$ given by\footnote{We
here introduce the nomenclature that subscripts of fixed points indicate the signs of
$u$, and hence also of $\Sigma_\varphi$ and $\varphi^\prime$, while superscripts
indicate the signs/values of $\bar{\varphi}$.}
\begin{equation} \label{kin.fixedpoint}
\mathrm{K}^+_{\pm}\!:\quad (\bar{\varphi}=1,\,u=\pm\sqrt2,\,v=1/\sqrt{3}),
\qquad \mathrm{K}^-_{\pm}\!:\,(\bar{\varphi}=-1,\,u=\pm\sqrt2,\,v=1/\sqrt{3}).
\end{equation}

The boundary set $v=1/\sqrt{3}$ may also contain fixed points on its intersection
with the exponential potential boundaries $\bar{\varphi} = \pm 1$, which we will
discuss in connection with these boundaries, and in its interior
$-1 <\bar{\varphi} < 1$. The latter occurs if the potential has one or several
extrema at some value(s) $\varphi = \varphi_0$ of the scalar field, which yields
$\lambda(\varphi_0)=0$, or if the potential is constant, {\it i.e.} if
$\lambda(\varphi)=0$, which yields a line of fixed points (see Figure~\ref{fig:3DLCDM}
in Section~\ref{sec:doubleexp}),
\begin{equation} \label{DS.fixed}
\mathrm{dS}^{\varphi_*}\!:\quad (\bar{\varphi}, u,v)
= (\bar\varphi_*, 0,1/\sqrt3) \quad \text{with} \quad
-1\leq \bar{\varphi}_*\leq 1.
\end{equation}
We will come back to this solvable case when discussing exponential potentials,
since $\lambda=0$ can be viewed as a special case of those models, but we
here note that $\mathrm{dS}^{\varphi_*}$ forms a global sink of these models.

A first classification of the models with varying $\lambda(\varphi)$ is given by
the number and character of extrema of the potential. For simplicity we consider
two special cases:
\begin{itemize}
\item[(i)] a monotonically decreasing potential with $\lambda(\varphi)>0$,
\item[(ii)] a potential with a single positive minimum at $\varphi=\varphi_0$ with
$\lambda(\varphi_0)=0$, $\lambda_{,\varphi}(\varphi_0)<0$, and $\lambda_-\geq 0$
and $\lambda_+ \leq 0$.
\end{itemize}
For case (ii) with $\lambda(\bar{\varphi}_0)=0$, $\lambda_{,\varphi}(\bar\varphi_0)<0$,
there is an isolated fixed point given by
\begin{equation} \label{DS0.fixed}
\mathrm{dS}^{0}\!:\quad (\bar\varphi, u,v)
=(\bar{\varphi}_0, 0,1/\sqrt3) \quad\text
{with}\quad \bar{\varphi}_0 \in (-1,1).
\end{equation}
It can be shown that the eigenvalues of $\mathrm{dS}^{0}$ have negative real
parts\footnote{\label{eigenvalues.dS0} The eigenvalues for the fixed point
$\mathrm{dS}^{0}$ in the $(\bar{\varphi},u,v)$ state space are
$-(3/2)[1\pm\sqrt{1+(4/3)\lambda_{,\varphi}(\varphi_0)}]$, where
$\lambda_{,\varphi}(\varphi_0) < 0$, and $-3$.} and hence
that $\mathrm{dS}^{0}$ is a local sink, which turns out to be global, and
if $\lambda_{,\varphi}(\bar{\varphi}_0)<-3/4$
then this sink is a spiral on the $v=1/\sqrt{3}$ ($\Omega_\varphi=1$)
boundary. Both $\mathrm{dS}^{\varphi_*}$ and $\mathrm{dS}^{0}$ are
characterized by $(u,v) = (0,1/\sqrt{3})$ which yields $w_\varphi=-1$, $q=-1$,
$\Omega_V=1$. Fixed points for which $(u,v) = (0,1/\sqrt{3})$ thereby
correspond to the de Sitter model and are therefore referred to as
\emph{de Sitter fixed points}.

\subsection{The exponential potential boundary sets $\bar\varphi=\pm1$}

Depending on the parameters
$\lambda_{\pm}\equiv \lim_{\varphi\rightarrow\pm\infty}\lambda(\varphi)$,
there are, in addition to the kinaton fixed points and the end points of the
lines of FL fixed points, three possible fixed points in each of the boundary sets
$\bar\varphi=\pm 1$: $\mathrm{dS}^{\pm}$, $\mathrm{P}^{\pm}$, ${\mathrm S}^{\pm}$.
To discuss the $\bar{\varphi}=\pm 1$ boundaries we for notational brevity
drop the subscript $\pm$ on $\lambda_\pm$ and the superscripts, which indicate
the signs of $\bar{\varphi}$ at the fixed points. For simplicity we
consider $\lambda\geq 0$ since the transformation
$\varphi\rightarrow -\varphi$ leads to $u\rightarrow - u$ and hence that
the fixed points and orbits for $\lambda<0$ are obtained from those with
$\lambda>0$ by reflecting these in $u=0$. The fixed points on the
$\bar{\varphi} = \pm 1$ boundaries (apart from the kinaton and FL
fixed points) can thereby be described as follows:
\begin{subequations} \label{fixed.points}
\begin{alignat}{3}
\mathrm{dS}&\!:\quad (u,v)\, &=& \, (0 , 1/\sqrt3),
&\quad\text{when}&\quad \lambda =0, \label{deS.pm} \\
\mathrm{P}&\!:\quad
(u,v)\, &=& \, (\lambda/\sqrt3 , 1/\sqrt3),
&\quad\text{when}&\quad 0<\lambda< \sqrt{6}, \label{S.pm} \\
{\mathrm S}&\!:\quad (u,v)\, &=& \,
(1,1/\lambda),
&\quad\text{when}&\quad \sqrt{3} < \lambda.
\label{Ss.fixedpoint}
\end{alignat}
\end{subequations}
The motivation for the notation of the kernels for the fixed points is as follows:
$\mathrm{P}$ corresponds to a power law
inflation solution when\footnote{It follows from~\eqref{qdef}
that $q=-1+\lambda^2/2$, and hence $\mathrm{P}$ corresponds to an accelerating
($q<0$) model when $0<\lambda<\sqrt{2}$, see Table~\ref{Table:fixed.points}.}
$0<\lambda<\sqrt{2}$, and $\mathrm{P}$ therefore stands for \emph{power law};
$\mathrm{dS}$ denotes \emph{de Sitter}, where $\mathrm{dS}$
corresponds to the limit of $\mathrm{P}$ when $\lambda\rightarrow 0$;
finally $\mathrm{S}$ corresponds to a scaling solution, and
$\mathrm{S}$ therefore represents \emph{scaling}.\footnote{The nomenclature scaling
arises  from the fact that $w_\varphi = w_\mathrm{m} = 0$ at $\mathrm{S}$ and
$\rho_\mathrm{m}(N)/\rho_\varphi(N) = \Omega_\mathrm{m}/\Omega_\varphi = -1 + \lambda^2/3$,
where $\rho_\mathrm{m} = \rho_{\mathrm{m},0}\exp(-3N)$, and
thus the scalar field mimics the dynamics of the fluid with a
constant ratio between both energy densities.}

\begin{figure}[ht!]
\begin{center}
\subfigure[$\lambda=0$]{\label{L0}
\includegraphics[width=0.35\textwidth]{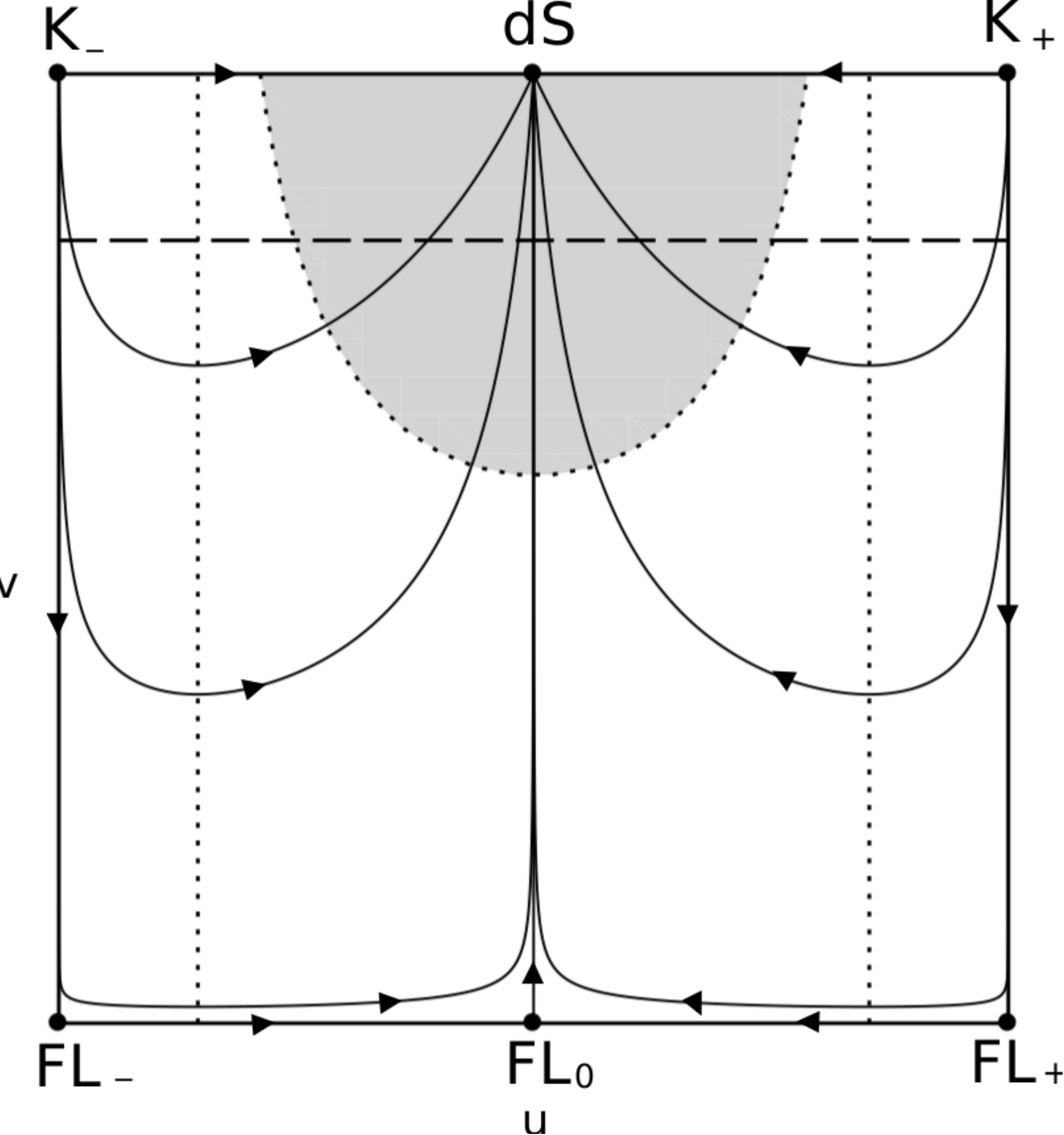}}
\hspace{1.5cm}
\subfigure[ $0<\lambda\leq\sqrt{3}$.]{\label{exp.pot1}
\includegraphics[width=0.35\textwidth]{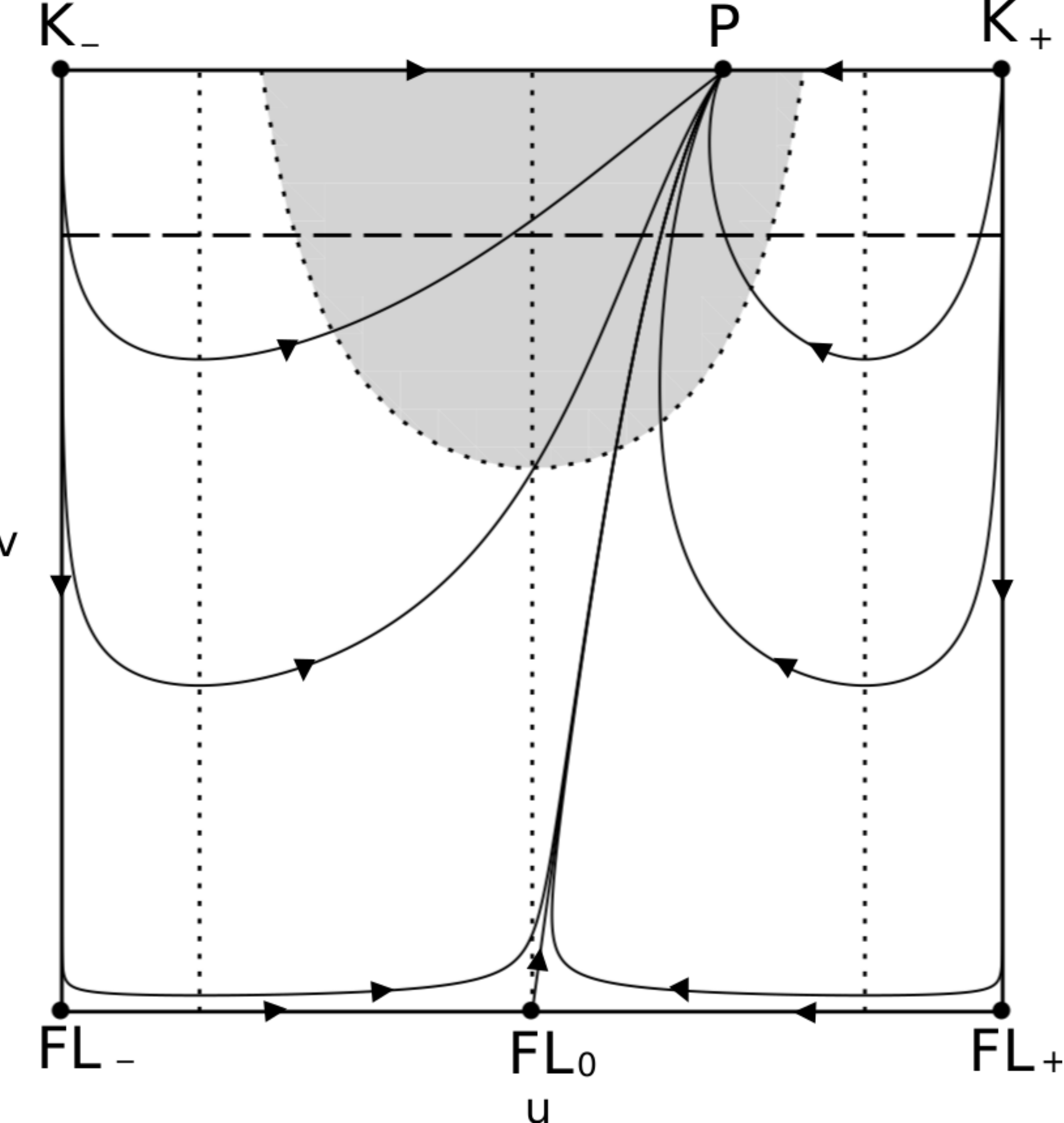}}\\
\subfigure[$\sqrt{3}<\lambda<\sqrt{6}$]{\label{exp.pot2}
\includegraphics[width=0.35\textwidth]{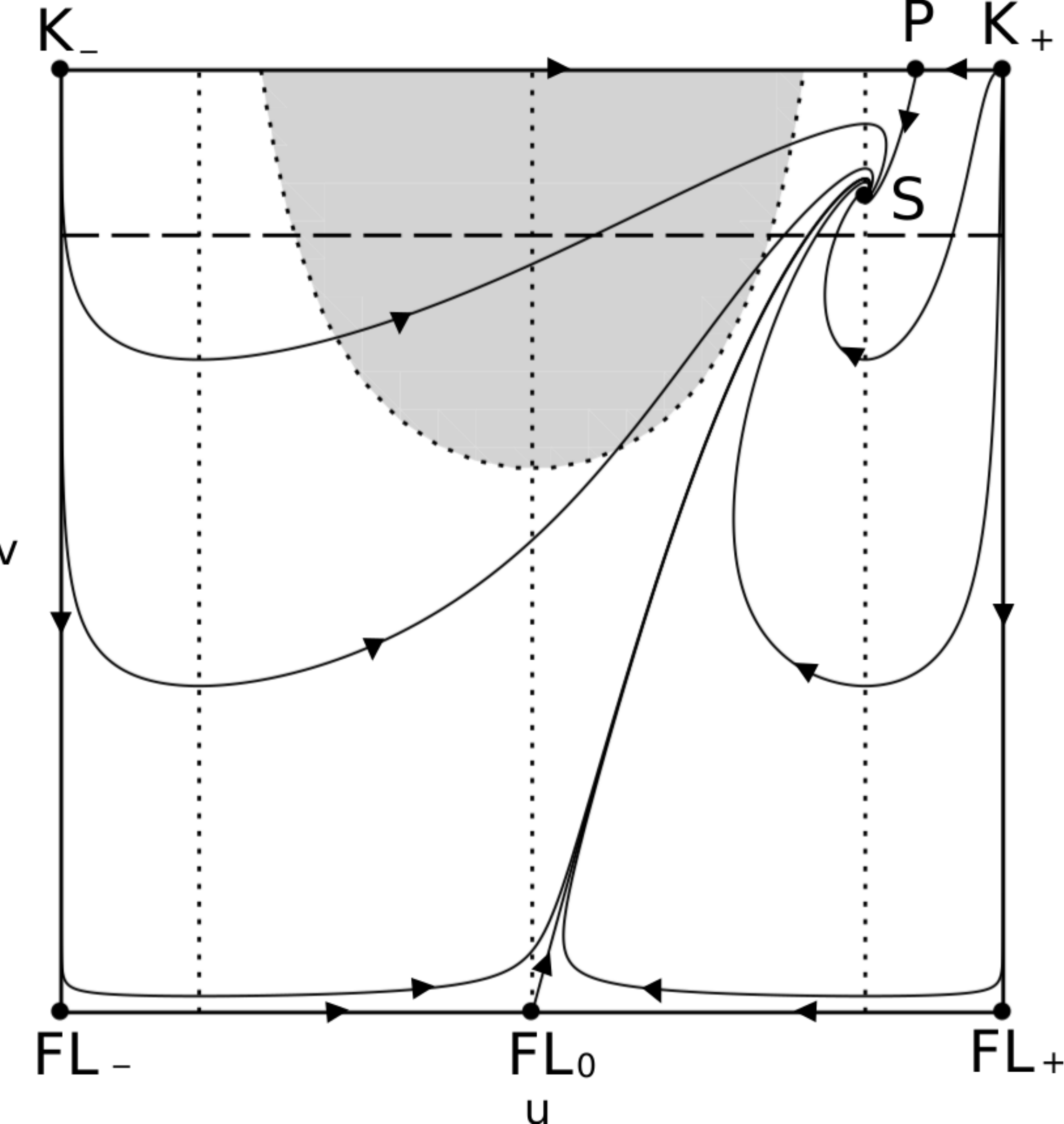}}
\hspace{1.5cm}
\subfigure[$\sqrt{6}\leq\lambda$]{\label{exp.pot3}
\includegraphics[width=0.35\textwidth]{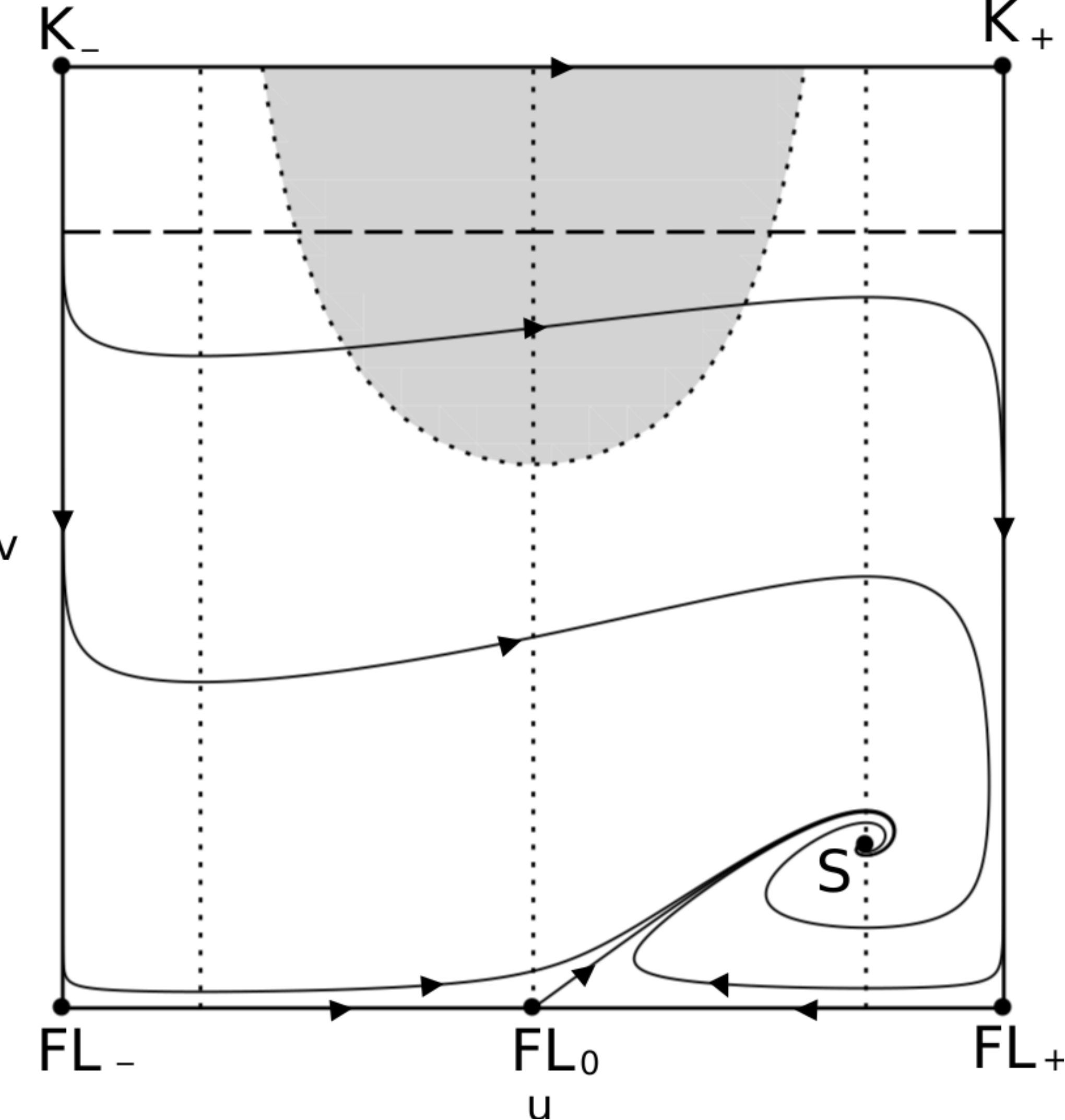}}
\vspace{-0.5cm}
\end{center}
\caption{Orbits for the $\lambda=\mathrm{constant}$ models that illustrate the orbit structures
on the boundary sets $\bar{\varphi}={\pm1}$ for the $\lambda=0$ case and the three parameter intervals
$0 < \lambda \leq \sqrt{3}$, $\sqrt{3} < \lambda < \sqrt{6}$, $\sqrt{6} \leq \lambda$
(with representative values $\lambda = 1, 2, 10$ in the figures). The shaded state space region
depicts $q<0$; note that if $\lambda < \sqrt{2}$ then $q<0$ for the fixed point $\mathrm{P}$. The horizontal dashed line indicated where $\Omega_{\varphi}=0.68.$
The dotted lines at $u=\pm1$ and $u=0$ indicate where $w_\varphi=0$, $v^\prime = 0$
and $w_\varphi=-1$, respectively. Between (outside)
the dotted lines $u=\pm1$ it follows that $-1 \leq w_\varphi<0$ ($w_\varphi>0$),
and that $\Omega_\varphi$ is increasing (decreasing).
}
\label{Fig:Box2d}
\end{figure}

As follows from the \emph{global} analysis of the $\lambda=\mathrm{constant}$ models given
in~\cite{alhetal22}, there are three qualitatively different orbit structures
that depend on the value of $\lambda$ for these models; specializing the global results
in~\cite{alhetal22} to dust as matter yields (for details, see~\cite{alhetal22})
\begin{equation}\label{bifurcationcases}
0 \leq \lambda \leq \sqrt{3}, \qquad \sqrt{3} < \lambda < \sqrt{6}, \qquad \sqrt{6} \leq \lambda,
\end{equation}
for which the different orbit structures are depicted in Figure~\ref{Fig:Box2d}.
Since the $\lambda=0$ models contain the $\Lambda$CDM model, given by the heteroclinic
separatrix orbit $\mathrm{FL}_0 \rightarrow \mathrm{dS}$ at $u=0$, and since the $\Lambda$CDM model
plays a special role for observational comparisons, the orbit structure for this case is
given separately, even though it is qualitatively the same as that for
$0<\lambda \leq \sqrt{3}$. As proven in~\cite{alhetal22}, all orbits
for models with constant $\lambda$ are either fixed points or heteroclinic orbits.

The fixed point $\mathrm{K}_-$ is a source in the $uv$-space, as is $\mathrm{K}_+$ when
$0\leq \lambda < \sqrt{6}$, but $\mathrm{K}_+$ becomes a saddle when $\sqrt{6} \leq \lambda$.
This is due to the bifurcation at $\lambda=\sqrt6$ where $\mathrm{P}$ leaves the state
space through $\mathrm{K}_+$, but $\mathrm{P}$ ($\mathrm{dS}$ when $\lambda=0$) is the future
attractor in the $uv$-space when $\lambda \leq\sqrt{3}$. The fixed point $\mathrm{S}$
enters the state space through $\mathrm{P}$ when $\lambda = \sqrt{3}$, yielding a bifurcation
that transfers the stability from $\mathrm{P}$ to $\mathrm{S}$ when $\lambda > \sqrt{3}$,
thereby leading to that $\mathrm{S}$ becomes the future attractor in the $uv$-space,
while $\mathrm{P}$ becomes a saddle from which the orbit $\mathrm{P}\rightarrow\mathrm{S}$
originates when $\sqrt{3} < \lambda < \sqrt{6}$.

For all values of $\lambda$, $\mathrm{FL}_0$ is a saddle from which one interior orbit originates.
In the case $\lambda=0$ this heteroclinic orbit, $\mathrm{FL}_0\rightarrow\mathrm{dS}$,
describes the $\Lambda$CDM solution. By changing $\lambda$ gradually from $\lambda=0$
the $\Lambda$CDM model is continuously deformed into the heteroclinic orbit
$\mathrm{FL}_0\rightarrow\mathrm{P}$, and to
$\mathrm{FL}_0\rightarrow\mathrm{S}$ when $\lambda > \sqrt{3}$. The orbit structures
for the different cases are depicted in Figure~\ref{Fig:Box2d}.
%

Finally, note that the fixed point $\mathrm{FL}$ in the $(\Sigma_\varphi,\Omega_\mathrm{m})$
state space at $(\Sigma_\varphi,\Omega_\mathrm{m}) = (0,1)$ is replaced by the boundary set
$v=0$ in the $(u,v)$ formulation, which consists of three fixed points,
$\mathrm{FL}_\pm$ and $\mathrm{FL}_0$, connected by the heteroclinic orbits
$\mathrm{FL}_\pm\rightarrow\mathrm{FL}_0$. Since $\mathrm{FL}_\pm$ are saddles
on the boundary, the two different formulations yield the same interior orbit structure,
cf. Figure 3 in~\cite{alhetal22}. The different fixed points in the box state space
$(\bar{\varphi},u,v)$ for the variable $\lambda(\varphi)$ in cases (i) and (ii) and their
values for $\Omega_\varphi$, $w_\varphi$ and $q$ are summarized in Table~\ref{Table:fixed.points}.
\begin{table}
\begin{center}
\begin{tabular}{|m{2.5cm}|m{2.5cm}|m{1cm}|m{2.0cm}|m{2.0cm}|}\hline
Name & $(u,v)$ & $\Omega_\varphi$ & $w_\varphi$ & $q$ \\ \hline
$\mathrm{K}^+_\pm,\,\mathrm{K}^-_\pm$ & $(\pm\sqrt{2},1/\sqrt{3})$ & $1$ & $1$ & $2$ \\ \hline
$\mathrm{dS}^\pm,\,\mathrm{dS}^0$  & $(0,1/\sqrt{3})$ & $1$ & $-1$ & $-1$ \\ \hline
$\mathrm{P}^\pm$ & $(\lambda_\pm,1)/\sqrt{3}$ & $1$ & $-1+\lambda_\pm^2/3$ & $-1+\lambda_\pm^2/2$ \\ \hline
$\mathrm{S}^\pm$ & $(1,1/\!\mid\!\lambda_\pm\!\mid)$ & $3/\lambda_\pm^2$ & $0$ & $1/2$ \\ \hline
$\mathrm{FL}_0^{\varphi_*}$ & $(0,0)$ & $0$ & $1$ & $1/2$ \\ \hline
$\mathrm{FL}_\pm^{\varphi_*}$ & $(\pm\sqrt{2},0)$ & $0$ & $1$ & $1/2$ \\ \hline
\end{tabular}
\caption{Summary of possible fixed points in the box state space $(\bar{\varphi},u,v)$
for (i) a monotonically decreasing potential and (ii) a potential with a single positive
minimum; their values for $u$ and $v$, while the signs/values of $\bar{\varphi}$
are indicated by the superscripts; their
values for $\Omega_\varphi$, $w_\varphi$ and the deceleration parameter $q$. The fixed points
$\mathrm{dS}^\pm$, $\mathrm{P}^\pm$, $\mathrm{S}^\pm$, only exist if $\lambda_\pm=0$,
$\lambda_\pm^2 < 6$, $\lambda_\pm^2 > 3$, respectively, while $\mathrm{dS}^0$ requires that
the potential has a positive minimum at some value $\varphi = \varphi_0$.
}\label{Table:fixed.points}
\end{center}
\end{table}
%
%
%

\subsection{Model classification and asymptotics\label{asymptotics}}

Recall that (apart from $\lambda=0$) we for simplicity consider two cases: (i) a monotonically
decreasing potential with $\lambda(\varphi)>0$; (ii) a potential with a single positive minimum
at $\varphi=\varphi_0$ such that $\lambda(\varphi_0) = 0$, $\lambda_{,\varphi}(\varphi_0)<0$
and $\lambda_-\geq 0$, $\lambda_+ \leq 0$.
Further subclassification is based on the combinatorial possibilities of the
three parameter intervals~\eqref{bifurcationcases} for $\lambda_-$ and $\lambda_+$.
As follows from the monotonic function $3H^2$ given by~\eqref{3H2mon},
local analysis of the fixed points, and the complete description of the
boundary orbit structures,\footnote{The asymptotics for the orbits on the $\bar{\varphi} = \pm 1$
boundaries were given in connection with the constant $\lambda$ models while the asymptotics for
the orbits on the $v=1/\sqrt{3}$ boundary are quite straightforward as well.}
this yield the different asymptotic possibilities and global orbit structures for
(i) and (ii) (in case (ii) $\mathrm{dS}^0$ is the future attractor for all interior
orbits, and also for the orbits on the scalar field dominated boundary $v = 1/\sqrt{3}$).

We will now describe the asymptotics for the
interior orbits, {\it i.e.} orbits for which  $-1 < \bar\varphi < 1$,
$-\sqrt{2} < u < \sqrt{2}$,\, $0 < v < 1/\sqrt{3}$, thereby describing solutions
with $\rho_\mathrm{m}>0$ and $V(\varphi)>0$, for the
two cases (i) and (ii):
\begin{itemize}
\item[(i)] \emph{Except} for a one-parameter set of heteroclinic orbits originating
from $\mathrm{FL}_0^{\varphi_*}$, where one orbit originates from each fixed point,
a one-parameter set of heteroclinic orbits from $\mathrm{P}^-$ when $\sqrt{3}<\lambda_-<\sqrt{6}$,
and one heteroclinic orbit from $\mathrm{S}^-$ when $\lambda_-> \sqrt{3}$, \emph{all}
interior orbits originate from the source $\mathrm{K}_-^+$,
and also from $\mathrm{K}_+^-$ when $0\leq \lambda_-<\sqrt{6}$.
\item[(ii)] \emph{Except} for a one-parameter set of heteroclinic orbits
originating from $\mathrm{FL}_0^{\varphi_*}$, where one orbit originates from each fixed
point, a one-parameter set of heteroclinic orbits from $\mathrm{P}^-$ ($\mathrm{P}^+$) when
$\sqrt{3}<\lambda_-<\sqrt{6}$ ($-\sqrt{6}<\lambda_+<-\sqrt{3}$),
and one heteroclinic orbit from $\mathrm{S}^-$ ($\mathrm{S}^+$) when $\lambda_-> \sqrt{3}$
(when $\lambda_-< -\sqrt{3}$), interior orbits originate from the heteroclinic cycle
$\mathrm{K}_-^-\rightarrow\mathrm{K}_+^-\rightarrow\mathrm{K}_+^+\rightarrow\mathrm{K}_-^+\rightarrow\mathrm{K}_-^-$
if $\lambda_-\geq \sqrt{6}$ \emph{and} $\lambda_+ \leq -\sqrt{6}$,\footnote{More precisely,
the heteroclinic cycle is their so-called $\alpha$-limit set.
A heteroclinic cycle is a closed heteroclinic chain, where the latter
consists of a concatenation of heteroclinic orbits, where the ending fixed point of one
heteroclinic orbit is the starting fixed point of the next one.} and otherwise
from $\mathrm{K}_-^+$, which is a source when $\lambda_-<\sqrt{6}$, or/and from $\mathrm{K}_+^-$,
which is a source when $\lambda_+ > -\sqrt{6}$.
\end{itemize}

Let us now for simplicity restrict the monotonic case (i) to $0 \leq \lambda_+<\sqrt{2}$,
since this yields future eternal acceleration. Then the future asymptotics for the interior
orbits for (i) and (ii) are as follows:
\begin{itemize}
\item[(i)] \emph{All} interior orbits end at the future attractor
${\cal A}^+ = \mathrm{P}^+$ (${\cal A}^+ = \mathrm{dS}^+$)
when $0 < \lambda_+<\sqrt{2}$ ($\lambda_+=0$).
\item[(ii)] \emph{All} interior orbits end at the future attractor ${\cal A}^+ = \mathrm{dS}^0$.
\end{itemize}

As a consequence of the above, all interior orbits
are heteroclinic orbits, as are all boundary orbits apart from the fixed points,
except in the case (ii) when $\lambda_- \geq \sqrt{6}$ \emph{and} $\lambda_+ \leq -\sqrt{6}$,
since the past attractor, ${\cal A}^-$, then is the above mentioned heteroclinic cycle.
All the above statements can be formally proved by using the monotonic function
in~\eqref{3H2mon}, the completely known structure of the boundary sets, notably
the global results in~\cite{alhetal22}, and the local analysis of the
fixed points, but for brevity we refrain from doing so here.

The above asymptotic properties lead to a useful result concerning $\Omega_\varphi$.
All interior orbits are future asymptotic to fixed points that satisfy $\Omega_\varphi=1$ and
they are all, except for the one-parameter set of unstable manifold orbits
of $\mathrm{FL}_0^{\varphi_*}$ and the single unstable manifold orbits of $\mathrm{S}^-$
and $\mathrm{S}^+$, past asymptotic to a fixed point
at $v=1/\sqrt{3}$, or the heteroclinic cycle at $v=1/\sqrt{3}$,
{\it i.e.} they originate from $\Omega_\varphi=1$. Any generic interior orbit hence attains
a positive minimum value of $\Omega_\varphi$, and thereby $v$, referred to as $v_\mathrm{min}$,
at some intermediate time, where it follows from~\eqref{alt.evol.eqs.2}
that this minimum occurs when $w_\varphi=0$, {\it i.e.} when $u=\pm 1$.
\section{Overview of bounded $\lambda(\varphi)$ quintessence\label{sec:overview}}

In this section we give an overview and classification of various types of quintessence
using the present $(\bar{\varphi},u,v)$ state space formulation. We also give some of
the key formulas for comparisons between the $\Lambda$CDM model and quintessence models,
noting that due to the observational success of the $\Lambda$CDM model, viable
quintessence models presumably cannot deviate too much from $\Lambda$CDM.

\subsection{The $\Lambda$CDM model\label{LCDMsec}}

\begin{figure}[ht!]
\begin{center}
\subfigure[$\Omega_\Lambda (N)$]{\label{LCDM_OL}
\includegraphics[width=0.35\textwidth]{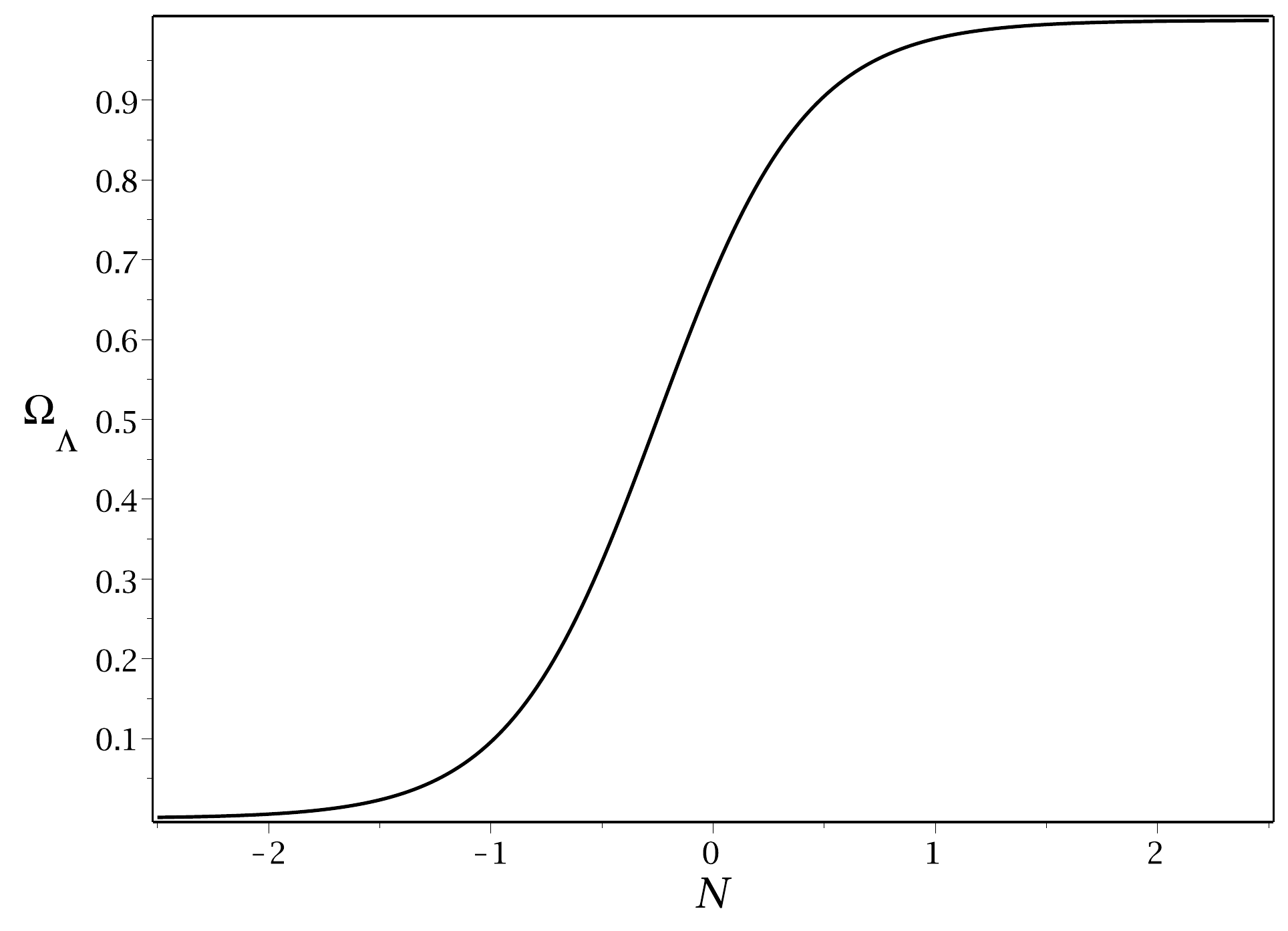}}
\hspace{1.5cm}
\subfigure[$H_{\Lambda}(N)/H_0$.]{\label{LCDM_E}
\includegraphics[width=0.35\textwidth]{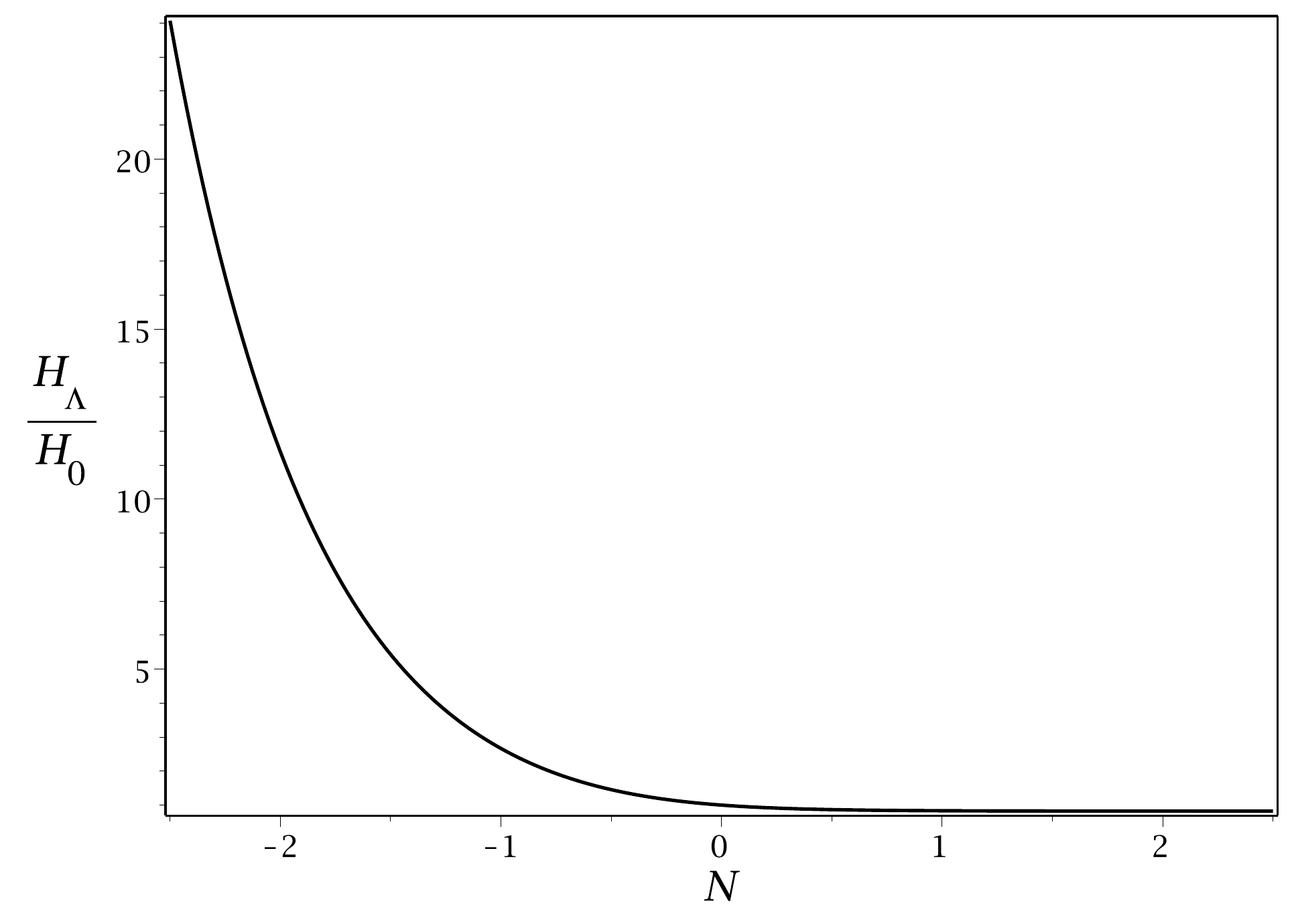}}
\vspace{-0.5cm}
\end{center}
\caption{$\Lambda$CDM graphs for $\Omega_\Lambda(N)$ and $H_\Lambda(N)/H_0$}
\label{Fig:LCDM}
\end{figure}

The $\Lambda$CDM model has the following key characteristics:
\begin{subequations}\label{LCDM}
\begin{align}
\Omega_\Lambda &= \frac{\Omega_{\Lambda,0}}{\Omega_{\Lambda,0}+(1-\Omega_{\Lambda,0})e^{-3N}},\label{OmL}\\
\left(\frac{H_\Lambda}{H_{\Lambda,0}}\right)^2 &= \left(\frac{\Omega_{\mathrm{m},0}}{\Omega_\mathrm{m}}\right)e^{-3N} =
\left(\frac{1-\Omega_{\Lambda,0}}{1-\Omega_\Lambda}\right)e^{-3N} = \Omega_{m,0}e^{-3N} + \Omega_{\Lambda,0},\\
q & = \frac12 - \frac32\Omega_\Lambda= - 1 + \frac32\Omega_\mathrm{m},\label{LCDMq}
\end{align}
\end{subequations}
where the expressions for $\Omega_\Lambda(N)$ and the Hubble variable $H_\Lambda(N)$ follow from spatial flatness,
$3H_\Lambda^2 = \rho_\mathrm{m} + \Lambda$, $\Omega_\mathrm{m} = 1 - \Omega_\Lambda$, and energy conservation
$\rho_\mathrm{m} = \rho_{\mathrm{m},0}\exp(-3N) = 3H_0^2\Omega_{\mathrm{m},0}\exp(-3N) =
3H_0^2(1 - \Omega_{\Lambda,0})\exp(-3N)$ (also, recall that $\Omega_{\Lambda,0} = \Lambda/3H_0^2$).
Setting $\Omega_{\Lambda,0} = 0.68$, it follows from~\eqref{LCDMq} that $N = -0.48$ at $q=0$. We
therefore expect that viable quintessence models begin accelerating when $N \gtrsim -0.5$,
$z \gtrsim 0.65$ (see Figure~\ref{Fig:LCDM}).\footnote{Recall that the redshift $z$ is defined as $z = \frac{a_0}{a} - 1 = e^{-N} - 1$
and hence that $N = - \ln(1+z)$.}

The basic criterion for quintessence evolution is that there is an early stage in
the universe in which the matter dominates the scalar field, $\Omega_\mathrm{m} \gtrsim 0.97$,
$\Omega_\varphi \lesssim 0.03$, followed by a decrease (increase) in $\Omega_\mathrm{m}$
($\Omega_\varphi$) to its present day value of approximately $0.32$ $(0.68)$. To obtain a
sense of at what time matter domination starts to decline and $\Lambda$ begins to effect
the evolution of the universe we can use equation~\eqref{OmL} to show that
\begin{equation} \label{matter.decline}
\Omega_\Lambda = 0.03 \quad \text {corresponds to}\quad N=-1.41,\,z=3.10,
\end{equation}
when $\Omega_{\Lambda,0}=0.68$, {\it i.e.} the effect of $\Lambda$ is felt only during
the last few $e$-fold(s) before the present time. The matter (dust) dominated epoch
is preceded by a radiation epoch. Since
$\rho_\mathrm{m}/\rho_\mathrm{rad} = \exp(N)\Omega_{\mathrm{m},0}/\Omega_{\mathrm{rad},0}
\approx \exp(N) 0.32/10^{-5}$, where thereby $\rho_\mathrm{m}=\rho_\mathrm{rad}$ when $N\approx - 10$
(the redshift $z=1100$ at decoupling corresponds to $N = -7$), which together with that matter
domination ends at $N\approx -1.4$ indicates that the matter (dust) dominated epoch is
$\Delta N \approx 8$ long. We expect similar results for observationally viable quintessence
models.
%

\subsection{Quintessence relations\label{viablequintsec}}

A similar calculation as for the $\Lambda$CDM model results in the following
quintessence relations:
\begin{subequations}\label{QCDM}
\begin{align}
\Omega_\varphi &= \frac{\Omega_{\varphi,0}}{\Omega_{\varphi,0}+(1-\Omega_{\varphi,0})e^{3\int_0^N w_\varphi(\tilde{N})d\tilde{N}}},\label{Omphiexpl}\\
\left(\frac{H_\varphi}{H_{\varphi,0}}\right)^2 &=
\left(\frac{1-\Omega_{\varphi,0}}{1-\Omega_\varphi}\right)e^{-3N} =
\Omega_{m,0}e^{-3N} + \Omega_{\varphi,0}e^{-3\int_0^Nu^2(\tilde{N})d\tilde{N}},
\label{Hphiexpl}\\
q &= \frac12 + \frac32w_\varphi\Omega_\varphi.
\end{align}
\end{subequations}
Equation~\eqref{Omphiexpl} shows that
$w_\varphi(N)$ (and thereby $u(N)$ since $w_\varphi= u^2 - 1$)
determines $\Omega_\varphi(N)$, which in turn yields $H_\varphi(N)$ and $q(N)$.

Quintessence evolution is defined to begin during matter dominance when
$\Omega_\mathrm{m}\approx 1$, and hence $\Omega_\varphi \approx 0$. The scalar
field $\varphi(N)$ is then effectively a test field that does not affect the
evolution of spacetime, as can be seen from~\eqref{Hphiexpl}, which yields
\begin{equation} \label{H/H0}
\left(\frac{H_\varphi}{H_{\varphi,0}}\right)^2 =
\left(\frac{\Omega_{\mathrm{m,0}}}{\Omega_\mathrm{m}}\right)e^{-3N} \approx \Omega_{\mathrm{m,0}}\,e^{-3N}.
\end{equation}
During matter dominance, for not too large $\lambda(\varphi)$, $w_\varphi=u^2-1$ is driven
toward $w_\varphi=-1$ ($u=0$) since $u'|_{v=0} = -\frac32(2-u^2)u$, but this behaviour has
no observable significance. After the early matter dominated epoch a quintessence model
should, like the $\Lambda$CDM model, have a monotonic increase of $\Omega_\varphi$, which
requires $u^2<1$, until $\Omega_\varphi > \Omega_\mathrm{m}$, where, eventually,
$\Omega_\varphi \rightarrow 1$, since $\Omega_\varphi=1$ at the future global sink
${\cal A}^+$.

Basset {\it et al.} (2008)~\cite{basetal08} state that nucleosynthesis yields the
bound $\Omega_\varphi< 0.034$, which also holds during decoupling in the
matter dominated epoch. We therefore divide evolution relevant for quintessence
as follows:
\begin{itemize}
\item a \emph{matter dominated epoch} during which
\begin{equation} \label{matter.epoch}
\Omega_\mathrm{m} \gtrsim 0.97,\quad
\Omega_\varphi \lesssim 0.03,
\end{equation}
and hence $v \lesssim 0.1$, where the scalar field is
approximately a test field, and
\item a \emph{quintessence epoch} where $\Omega_\varphi \gtrsim 0.03$
and $v \gtrsim 0.1$ while
$\Omega_\mathrm{m} \lesssim 0.97$. This epoch is subdivided into an
\emph{observable quintessence epoch} from when $\Omega_\varphi \approx 0.03$
to the present where $\Omega_\varphi = 0.68$, and a \emph{future quintessence epoch} when
$\Omega_\varphi$ evolves from $\Omega_\varphi = 0.68$ to $\Omega_\varphi = 1$.\footnote{The
details of the bounds for (significant) quintessence (content) evolution will change as
observations become increasingly accurate. This, however, will not affect
our qualitative conclusions, which are robust under such changes.}
\end{itemize}
At the end of section~\ref{asymptotics} we showed
that on a generic interior orbit the variable $v$ attains a positive
minimum value $v_\mathrm{min}$ at some intermediate time.
Thus a necessary condition for a generic interior orbit to have a matter dominated
epoch with $v \lesssim0.1$  is that $v_\mathrm{min}<0.1$.

Since the quintessence evolution of $\Omega_\varphi(N)$ and $H_\varphi(N)$ for
observationally viable quintessence models presumably differs by less than
$10\%$ when compared to $\Lambda$CDM, equation~\eqref{matter.decline} suggests
that the observational quintessence epoch takes place between $N\approx -1.5$ ($z\approx 3.5$)
and $N=0$. Furthermore, as discussed earlier, the quintessence epoch is preceded by a
matter (dust) dominated epoch with $\Delta N\approx 8$. Moreover, combining the matter
dominated epoch with a preceding radiation dominated epoch, here neglected, leads to
$\Delta N > 8$. From a dynamical systems perspective radiation and dust structurally are
quite similar and we will therefore for simplicity in this paper replace radiation with
dust and assume that $\Delta N > 8$ during the matter dominated epoch (the role of
inflation preceding the radiation dominated epoch will be dealt with in a future
paper~\cite{alhugg22}). Such a long matter dominated $e$-fold interval \emph{severely
restricts viable initial data} and is \emph{only possible for orbits that come
extremely close to one or several of the fixed points} $\mathrm{FL}_0^{\varphi_*}$,
$\mathrm{FL}_\pm^{\varphi_*}$ or $\mathrm{S}^\pm$ (when $v(\mathrm{S}^\pm)< 0.1$).

\subsection{$\Lambda$CDM and quintessence comparisons\label{comparisons}}

In the introduction we noted that the $\Lambda$CDM model can be thought of
as a special quintessence model with constant potential $V=\Lambda$ and constant
scalar field and equation of state parameter $w_\Lambda=-1$. In order to compare
a quintessence model with the  $\Lambda$CDM model we identify the models at the
present time as regards rate of expansion and matter content. Specifically we
require that
\begin{equation}
H_{\varphi,0} = H_{\Lambda,0} = H_0,\qquad
\Omega_{\Lambda,0} = \Omega_{\varphi,0}=1-\Omega_{m,0},
\end{equation}
where $H_0$ and $\Omega_{m,0}$ are the observed Hubble parameter and
the dimensionless Hubble-normalized matter density at the present time.

From~\eqref{LCDM} and~\eqref{QCDM} it follows that
\begin{subequations}\label{Hcomp}
\begin{align}
\left(\frac{H_\varphi}{H_\Lambda}\right)^2 &=
\frac{1-\Omega_\Lambda(N)}{1-\Omega_\varphi(N)},\\
\frac{1}{\Omega_\varphi} - \frac{1}{\Omega_\Lambda} &=
\left(\frac{1 - \Omega_{\Lambda,0}}{\Omega_{\Lambda,0}}\right)\left(e^{3\int_0^Nu^2(\tilde{N})d\tilde{N}} - 1\right)e^{-3N},
\end{align}
\end{subequations}
which leads to the inequalities
\begin{subequations}\label{Hineq}
\begin{alignat}{3}
H_\varphi(N) &> H_\Lambda(N), &\qquad \Omega_\varphi(N) &> \Omega_\Lambda(N), \quad &\text{when}\quad N &< 0,\\
H_\varphi(N) &< H_\Lambda(N), &\qquad \Omega_\varphi(N) &< \Omega_\Lambda(N), \quad &\text{when}\quad N &> 0,
\end{alignat}
\end{subequations}
where the inequalities for the Hubble variable also follow from
\begin{equation}
\left(\frac{H_\varphi}{H_0}\right)^2 - \left(\frac{H_\Lambda}{H_0}\right)^2
= \Omega_{\Lambda,0}\left(e^{-3\int_0^N u^2(\tilde{N})d\tilde{N}} - 1\right).
\end{equation}
For the present models, which comprise the two cases (i) and (ii), introduced earlier
following eq.~\eqref{DS.fixed}, there are several different types of quintessence,
which we now characterize, beginning with (i).

\subsection{Monotonic potentials}\label{sec:monotonic}

In this subsection we consider models with a monotonically decreasing potential
(case (i)) with $\lambda_+<\sqrt2$. We first note that for interior orbits
(as well as for interior orbits on the $\Omega_\varphi = 1$, $v=1/\sqrt{3}$ boundary)
\begin{equation}
u^\prime|_{u=0} = 2v\lambda(\bar{\varphi}) > 0,
\end{equation}
as follows from~\eqref{u.prime}. Hence $u=0$ acts as a semi-permeable membrane
for the dynamics, which together with the monotonic function and local fixed point
analysis implies that orbits with $u<0$ initial data eventually enters the invariant
$u>0$ part of the state space. This corresponds to that a scalar field that is initially
moving toward the potential `wall' eventually bounce against it (recall that
$u \propto \varphi^\prime/\sqrt{\Omega_\varphi}$ due to eq.~\eqref{tent.to.box.map}).

\subsubsection*{Thawing and freezing quintessence \label{thaw.freeze}}

Caldwell and Linder (2005)~\cite{callin05} defined \emph{thawing} as
$w_\varphi^\prime>0$ when $w_\varphi\approx - 1$ and \emph{freezing} as
$w_\varphi^\prime>0$ when $w_\varphi>-1$. This motivates defining a
\emph{thawing quintessence model} as a model for which $w_\varphi\approx - 1$ and
$w_\varphi^\prime >0$ at the beginning of the observational quintessence epoch,
which commences when $\Omega_\varphi \approx 0.03$ at some $e$-fold time
$N=N_\mathrm{quint}$. In analogy, a \emph{freezing quintessence model} is
characterized by $w_\varphi> - 1$ and $w_\varphi^\prime <0$ at $N_\mathrm{quint}$.
Note that the thawing (freezing) property may change sometime later during the
quintessence epoch.

Thawing and freezing quintessence models correspond to orbits that are schematically
described as follows:
%
\begin{itemize}
\item[(U)] a one parameter family of heteroclinic `U-orbits' 
$\mathrm{FL}_0^{\varphi_*} \rightarrow {\cal A}^+$, {\it i.e.}
the unstable manifolds of the fixed points of
$\mathrm{FL}_0^{\varphi_*}$ that join a fixed
point $\mathrm{FL}_0^{\varphi_*}$ with $-1<\bar\varphi_*<1$ to the future attractor
${\cal A}^+$ (the fixed point $\mathrm{dS}^+$ when $\lambda_+=0$; 
$\mathrm{P}^+$ when $0 <\lambda_+ < \sqrt{2}$), where U stands for  `unstable'.
\item[(S)] an open set of `S-orbits'
\begin{equation}\label{thafree}
\underbrace{{\cal A}^- \quad\longrightarrow\qquad}_{\text{pre-matter dominance}}\,
\underbrace{\mathrm{FL}_\pm^{\varphi_*} \quad\longrightarrow\qquad}_{\text{matter dominated frozen scalar field}}\,
\underbrace{\mathrm{FL}_0^{\varphi_*}\quad\longrightarrow\quad {\cal A}^+}_{\text{quintessence dynamics}}
\end{equation}
\end{itemize}
that shadow the U-orbits during intermediate and late times, where S stands
for `shadowing'.
The past attractor ${\cal A}^- $ depends on $\lambda_-$, which complicates
the details of the first step. When $\lambda_-< \sqrt{6}$ (for which
${\cal A}^- = \mathrm{K}_-^+\cup\mathrm{K}_+^-$) the first step is
$\mathrm{K}_-^+\rightarrow \mathrm{FL}_-^{\varphi_*}$ or
$\mathrm{K}_+^-\rightarrow \mathrm{FL}_+^{\varphi_*}$;
when $\lambda_-\geq \sqrt{6}$ (for which ${\cal A}^- = \mathrm{K}_-^+$) the
first case still holds, but the second is replaced by the sequence
$\mathrm{K}_-^+\rightarrow \mathrm{K}_-^-\rightarrow \mathrm{K}_+^-\rightarrow \mathrm{FL}_+^{\varphi_*}$.\footnote{Which
of the two routes,
$\mathrm{K}_-^+\rightarrow \mathrm{FL}_-^{\varphi_*} \rightarrow \mathrm{FL}_0^{\varphi_*} \rightarrow {\cal A}^+$ and $\mathrm{K}_+^-\rightarrow \mathrm{K}_-^-\rightarrow \mathrm{K}_+^-\rightarrow \mathrm{FL}_+^{\varphi_*} \rightarrow \mathrm{FL}_0^{\varphi_*} \rightarrow {\cal A}^+$
that is taken can be understood heuristically. The first case corresponds to that there is sufficient
early matter content to create enough early matter dominated friction so that the scalar field (almost) freezes
to a constant value before a `soft’/slow scalar field bounce in the matter dominated regime; the second case
corresponds to that there is sufficient kinetic scalar field content to produce a `sharp’/fast scalar field
bounce during an early scalar field dominated stage, {\it i.e.}, the route is determined by the past
ratio between the (kinetic) scalar field energy density and the matter energy density (obtainable
from the linearization at $\mathrm{K}_-^+$).}

The number of $e$-folds $\Delta N$ spent by S-orbits during the matter-dominated epoch,
where $\Omega_\mathrm{m} \gtrsim 0.97$, $\Omega_\varphi \lesssim 0.03$ and hence
$v \lesssim 0.1$, is highly dependent on the minimum value of $v=v_\mathrm{min}$ at
$u=\pm 1$; the smaller $v_\mathrm{min}$ the more $e$-folds, since this implies that
S-orbits come closer and thereby stay longer near the $\mathrm{FL}$ fixed points.
As discussed, viable models must spend $\Delta N> 8$ $e$-folds during the matter
dominated epoch. This forces $v=v_\mathrm{min}$ to be very small, and hence
an S-orbit will intermediately shadow
$\mathrm{FL}_+^{\varphi_*} \rightarrow\mathrm{FL}_0^{\varphi_*}$
or $\mathrm{FL}_-^{\varphi_*} \rightarrow\mathrm{FL}_0^{\varphi_*}$
very closely. This is subsequently followed by shadowing of the U-orbit
$\mathrm{FL}_0^{\varphi_*}\rightarrow{\cal A}^+$, where shadowing is further
strengthened by the fact that all interior orbits end at the future attractor
${\cal A}^+$. As a consequence the U-orbits describe the quintessence epoch
of the S-orbits extremely well. Before shadowing
$\mathrm{FL}_\pm^{\varphi_*} \rightarrow\mathrm{FL}_0^{\varphi_*}$, with frozen
$\bar{\varphi} \approx \bar{\varphi}_*$,
the S-orbits shadow orbits on the boundaries $u = \pm \sqrt{2}$ (the stable manifolds of the
fixed points $\mathrm{FL}_\pm^{\varphi_*}$, see Figure~\ref{BoxSS2}) very closely,
since the small value of $v=v_\mathrm{min}$ results in coming very close to $\mathrm{FL}_\pm^{\varphi_*}$.

We can draw some general conclusions about the models that are described by the
S-orbits. Recall that, before the quintessence epoch, 
S-orbits shadow orbits on the boundaries $u = \pm\sqrt{2}$,
where $w_\varphi = 1$, followed by shadowing of 
$\mathrm{FL}_\pm^{\varphi_*}\rightarrow \mathrm{FL}_0^{\varphi_*}$,
where $w_\varphi = -1$ at $\mathrm{FL}_0^{\varphi_*}$. Hence the quintessence 
epoch for S-orbits is characterized by a preceding stage where $w_\varphi(N)$ 
is \emph{approximated by a step function that steps down from $+1$ to $-1$};
the time of the rapid drop in $w_\varphi$ is determined by how closely
$\mathrm{FL}_\pm^{\varphi_*}\rightarrow \mathrm{FL}_0^{\varphi_*}$ is
shadowed ({\it i.e.} it is determined by $v_\mathrm{min}$), illustrated by
the discussion in connection with Figure~\ref{Figb_wDE} in the next section.

The distinction between thawing and freezing quintessence models is due to the
behaviour of $w_\varphi$ along the U-orbits
$\mathrm{FL}_0^{\varphi_*} \rightarrow {\cal A}^+$. We first consider the case
$\lambda_{+}=0$ so that ${\cal A}^+ = \mathrm{dS}^+$. Then $w_\varphi=-1$ at both
endpoints of the U-orbits, which implies that along each orbit $w_\varphi(N)$
\emph{must attain at least one local maximum} (colloquially, a `bump'), where
$w_\varphi^\prime <0$ changes sign from positive (thawing) to negative (freezing).
The details depend on the value of $\lambda_* \equiv \lambda(\varphi_*)$.
Loosely speaking, for small $\lambda_*$, {\it i.e.} $\lambda_* = {\cal O}(1)$,
there is a single bump of small amplitude occurring \emph{after} the
beginning of the quintessence epoch at $N_\mathrm{quint}$,
leading to a thawing quintessence model ($w_\varphi^\prime >0$ at
$N=N_\mathrm{quint}$), illustrated by Figures~\ref{Figb_wDE} and~\ref{Figc_wDE100}, 
with $\lambda_* = 1,2$, in the next section.
As $\lambda_*$ increases the amplitude of the bump increases and it occurs earlier,
eventually \emph{before} $N_\mathrm{quint}$, leading to a freezing quintessence model
($w_\varphi^\prime <0$ at $N=N_\mathrm{quint}$), illustrated by $\lambda_* = 9.9$
in Figure~\ref{Figc_wDE100} in the next section. Further increase of $\lambda_*$
eventually leads to scaling freezing quintessence (see below).

In the case $\lambda_{+}>0$ ${\cal A}^+ = \mathrm{dS}^+$ is replaced
with ${\cal A}^+ = \mathrm{P}^+$, where $w_\varphi=-1+\lambda_{+}^2/3$,
which leads to a plateau in the graph of $w_\varphi$ as $N\rightarrow\infty$ for each
U-orbit. When $\lambda_+ = O(1)$ this leads to greater variability in the graph $w_\varphi(N)$
in the observational quintessence epoch, {\it e.g.} for some values of $\lambda_*$  
a bump does not occur, as exemplified in Figure~\ref{Figc_wDE101} below.


\subsubsection*{Scaling freezing quintessence \label{scale.freeze}}

\emph{Scaling freezing quintessence} occurs for monotonically decreasing potentials (i)
with $\lambda_-\gtrsim 10$ and is characterized by an approximate scaling behaviour
initially, {\it i.e.} $w_\varphi \approx 0$ and
$\rho_\varphi(N) \propto \rho_\mathrm{m}\propto \exp(-3N)$ and hence
$\Omega_\varphi(N) \propto \Omega_\mathrm{m}(N)$, for at least some $e$-folds in the
matter dominated regime ($\Omega_\varphi(N) \lesssim 0.03$) followed by freezing
$w_\varphi'(N)<0$ into the quintessence epoch. Scaling freezing quintessence models are
associated with the unstable manifold of the scaling fixed point $\mathrm{S}^-$,
which we recall is given by $(u,v, \bar\varphi )=(1,1/\lambda_-,-1)$ with $w_\varphi=0$,
$\Omega_\varphi=3/\lambda_-^2$. Since the (freezing) quintessence epoch must be
preceded by a matter dominated scaling epoch it follows that the fixed point
$\mathrm{S}^-$ \emph{must be located in the matter dominated region of state space}
($\Omega_\varphi(N) \lesssim 0.03$, see~\eqref{matter.epoch}),
which requires $\lambda_- \gtrsim 10$.

Scaling freezing quintessence models are described by orbits of the following types:
\begin{itemize}
\item[(U)] a single heteroclinic \emph{scaling freezing orbit} $\mathrm{S}^- \rightarrow {\cal A}^+$
(the unstable manifold of $\mathrm{S}^-$) that joins the scaling fixed point
$\mathrm{S}^-$ to the future attractor ${\cal A}^+$ 
(the fixed point $\mathrm{dS}^+$ when $\lambda_+=0$; $\mathrm{P}^+$ when $0 <\lambda_+ < \sqrt{2}$),
\item[(S)] an open set of heteroclinic shadowing orbits that come very close to
$\mathrm{S}^-$
during an intermediate stage of their evolution and afterward
shadow the scaling freezing orbit $\mathrm{S}^- \rightarrow {\cal A}^+$:
\begin{equation}
\underbrace{\mathrm{K}_-^+ \longrightarrow\mathrm{K}_-^-\longrightarrow\cdots}_{\text{prescaling dynamics}}\,
\underbrace{\mathrm{S}^-\quad\longrightarrow\qquad{\cal A}^+}_{\text{scaling freezing quintessence}},
\end{equation}
\end{itemize}
where the separate stages are illustrated in Figures~\ref{fig:3Dm10p0}
and~\ref{fig:3Dm10p1}.\footnote
{There are similarities and differences
between scaling orbit attraction and the attractor solution/orbit in inflationary cosmology. In both cases
there is a stable manifold of co-dimension one of an isolated fixed point and a one-dimensional
unstable manifold, the `attractor' orbit. However, in contrast to the scaling orbit, which corresponds
to a positive eigenvalue, the inflationary attractor solution corresponds to a zero eigenvalue and
is thereby a center manifold orbit. The zero eigenvalue leads to that (A) the inflationary
attractor orbit attracts nearby orbits much more strongly than the scaling orbit, (B) orbits stay
much longer close to the de Sitter fixed point in a quasi-de Sitter stage than orbits stay close
to $\mathrm{S}^-$.} The $\cdots$ refers to one of the orbits in the boundary set $\bar\varphi=-1$
joining $\mathrm{K}_-^-$ to $\mathrm{S}^-$,
as shown in Figure~\ref{exp.pot3}. We note that $\mathrm{K}_-^-$ is the source and
$\mathrm{S}^-$ is the sink for orbits \emph{in} this boundary set, which facilitates the
shadowing by the orbits (S). It is important to note that to obtain orbits for which
the scaling property holds for several $e$-folds in the matter dominated regime we
have to ensure that the orbits (S) come extremely close to the scaling fixed point
$\mathrm{S}^-$. We will achieve this in the numerical simulations in
sections~\ref{sec:Case2} and~\ref{sec:Case3} by choosing appropriate initial values.
%


\subsection{Scaling oscillatory and oscillatory quintessence \label{sec:scale.osc}}

We now consider case (ii), {\it i.e.} a potential with a single positive minimum at
$\bar\varphi_0$ that gives rise to the de Sitter fixed point
$\mathrm{dS}^{0}\!\!: (\bar\varphi, u,v)=(\bar{\varphi}_0, 0,1/\sqrt3)$ as the future attractor.
The unstable manifold of $\mathrm{FL}_0^{\varphi_0}$, which is the straight line
$\bar\varphi=\bar\varphi_0$, $u=0$ terminating at $\mathrm{dS}^0$ ({\it i.e.} the heteroclinic
orbit $\mathrm{FL}_0^{\varphi_0}\rightarrow\mathrm{dS}^0$), corresponds to the $\Lambda$CDM model.
For brevity we consider the case when $\lambda_{,\varphi}(\bar{\varphi}_0) <-3/4$
so that $\mathrm{dS}^0$ is an attracting spiral, which causes orbits to spiral around the
straight $\Lambda$CDM orbit as they approach $\mathrm{dS}^0$. Hence for non-$\Lambda$CDM
orbits $\bar{\varphi}$ oscillates around $\bar\varphi_0$ and $u$
oscillates around $0$ while $v$ increases to $1/\sqrt3$ as $N\rightarrow\infty$, which
implies that $1+w_\varphi(N)$ undergoes continuing oscillations with damped amplitude
as the future attractor $\mathrm{dS}^0$ is approached. Thus, in the observable
quintessence epoch there are a finite number of oscillations of $1+w_\varphi(N)$ with
changes in the sign of $w_\varphi^\prime(N)$, {\it i.e.} the evolution successively changes
between thawing and freezing.

There are numerous pre-quintessence possibilities, depending on the values of
$\lambda_-$ and $\lambda_+$. As an example we will consider $\lambda_-\gg1$
and $-\lambda_+\gg1$, which means that the scaling fixed points ${\mathrm S}^-$ and ${\mathrm S}^+$
come into play. The unstable manifolds of ${\mathrm S}^{\pm}$ spiral around the straight line
orbit $\mathrm{FL}_0^{\varphi_0} \rightarrow \mathrm{dS}^0$ as they approach $\mathrm{dS}^{0}$.
We divide quintessence orbits for models with $\lambda_-\gg1$ and $-\lambda_+\gg1$ into two classes:
\begin{itemize}
\item \emph{Scaling oscillatory quintessence} is described by the scaling orbits
$\mathrm{S}^-\rightarrow\mathrm{dS}^0$ and $\mathrm{S}^+\rightarrow\mathrm{dS}^0$,
and orbits that come extremely close to $\mathrm{S}^-$ or $\mathrm{S}^+$. These latter orbits
subsequently shadow the scaling orbits and exhibit the scaling property during part of
the matter dominated epoch, followed by oscillations in the quintessence epoch.
\item \emph{Oscillatory quintessence} is described by $\mathrm{FL}_0^{\varphi_*}\rightarrow\mathrm{dS}^0$
orbits, and orbits that shadow these orbits, not coming extremely close to
$\mathrm{S}^-$ or $\mathrm{S}^+$, thereby not exhibiting the scaling property during part of
the matter dominated epoch, but they still undergo oscillations in the quintessence epoch.
\end{itemize}
For examples, see Figure~\ref{fig:3DOsc}; note, however, that generic orbits originate
from ${\cal A}^-$, which for these models is the heteroclinic cycle on the boundary of
the $\Omega_\varphi = 1$ boundary.

\section{Example: The double-exponential potential\label{sec:doubleexp}}

In this section we will illustrate some aspects of the previous general
discussion about quintessence using special cases of a simple example:
the double-exponential potential,
\begin{equation} \label{two.exp.pot}
V = M_-^4e^{-\lambda_-\varphi}+ M_+^4e^{-\lambda_+\varphi},\qquad
M_\pm >0,\qquad  \lambda_-> 0, \qquad \lambda_->\lambda_+.
\end{equation}
We will represent the orbits using figures in the box state space, augmented with
graphs of $w_\varphi(N)$ and $H_\varphi(N)/H_\Lambda(N)$.\footnote{The
double-exponential potential models have a fairly lengthy history in the literature.
An early paper is Barreiro {\it et al.} (2000)~\cite{baretal00}; see also, {\it e.g.},
Barro Calvo and Maroto (2006)~\cite{barmar06}, section IIB, Bassett {\it et al.}
(2008)~\cite{basetal08}, section 3.2 and Dunega {\it et al.} (2013)~\cite{dunetal13}, eq. (A.3).
In particular~\cite{basetal08} have given simulations of $w_\varphi(N)$ for initial data that
in effect yields the $\mathrm{S}^-$
scaling orbit using $\lambda_- = 9.43$ and several positive values of $\lambda_+$ between $0$
and $1$, as well as negative values between $0$ and $-30$, see their Figure 2, lower left panel;
note that the temporal range they use is $z\in[0,10]$, which corresponds to $N\in [-2.4,0]$.
}
To include the potential~\eqref{two.exp.pot} in the present framework we introduce
the bounded variable
\begin{equation}
\bar{\varphi} = \tanh(C\varphi + D), \qquad C = \frac12(\lambda_- -\lambda_+),\qquad
D = 2\ln(M_+/M_-).
\end{equation}
It follows that $\lambda$ is a linear function of $\bar{\varphi}$ given by
\begin{equation} \label{lambda.bar.phi}
\lambda(\bar{\varphi}) = \frac12\lambda_+(1 + \bar{\varphi}) +
\frac12\lambda_-(1 - \bar{\varphi}).
\end{equation}

Before beginning, for the purpose of comparison, we represent the $\Lambda$CDM model
by orbits in the box state space. We have noted that the $\Lambda$CDM model can be
viewed as a limiting quintessence model with constant potential $V(\varphi)=\Lambda$
and a constant scalar field, which results in $w_{\varphi}=-1$ ($u=0$). Since $\lambda=0$
it follows from~\eqref{u.prime} that the set $u=0$ is an invariant set and that the
orbits that represent the $\Lambda$CDM model are straight lines
$u=0,\bar\varphi=\mathrm{constant}$. Figure~\ref{fig:3DLCDM} shows this
one-parameter set of $\Lambda$CDM orbits which join the
$\mathrm{FL}_0^{\varphi_*}$ fixed points to the  de Sitter fixed points
$\mathrm {dS}^{\varphi_*}$. We will see that the invariant set $u=0$ in
the constant potential case, and the $\Lambda$CDM orbits that it contains,
becomes deformed for varying potentials.

We will use the following four illustrative choices of the parameters $\lambda_{\pm}$:
\begin{itemize}
\item[1)] the monotonic potential with $\lambda_-=1$, $\lambda_+=0$;
\item[2)] the monotonic potential with $\lambda_-=10$, $\lambda_+=0$;
\item[3)] the monotonic potential with $\lambda_-=10$, $\lambda_+=1$;
\item[4)] a potential with a positive minimum for which
$\lambda_-=20$, $\lambda_+=-10$ with $\mathrm{dS}^0$, being a stable spiral since
$\lambda_{,\varphi}(\varphi_0)=\lambda_{+}\lambda_{-}=-200<-3/4$, where
$\bar{\varphi}_0=1/3$, since~\eqref{lambda.bar.phi} results in
\begin{equation} \label{lambda.zero}
\lambda(\bar{\varphi}_0) = 0\quad\Rightarrow\quad
\bar{\varphi}_0= \frac{\lambda_{-} +\lambda_{+} }{\lambda_{-} -\lambda_{+}}.
\end{equation}
\end{itemize}
%

\subsection{Case 1: $\lambda_-=1$, $\lambda_+=0$ \label{sec:Case1}}

This case illustrates the box state space for the double exponential potential,
augmenting Figure~\ref{BoxSS2} with information that depends on the values of
$\lambda_\pm$, namely the fixed points
${\mathrm P}^-$ and $\mathrm {dS}^+$ and the stability of the kinaton fixed points
indicated by the arrows on the orbits joining them.  The details of the
orbits in the boundary sets $\bar\varphi =\pm1$ are given in Figures~\ref{L0}
and~\ref{exp.pot1}, respectively. Figure~\ref{BoxSS2} also shows the orbits in the
boundary sets $u=\pm\sqrt2$ and in the base of the box $v=0$.
\begin{figure}[ht!]     
\begin{center}
\subfigure[$\Lambda$CDM orbits in the box state space for a constant potential and hence
$\lambda=0$]{\label{fig:3DLCDM}
\includegraphics[width=0.45\textwidth]{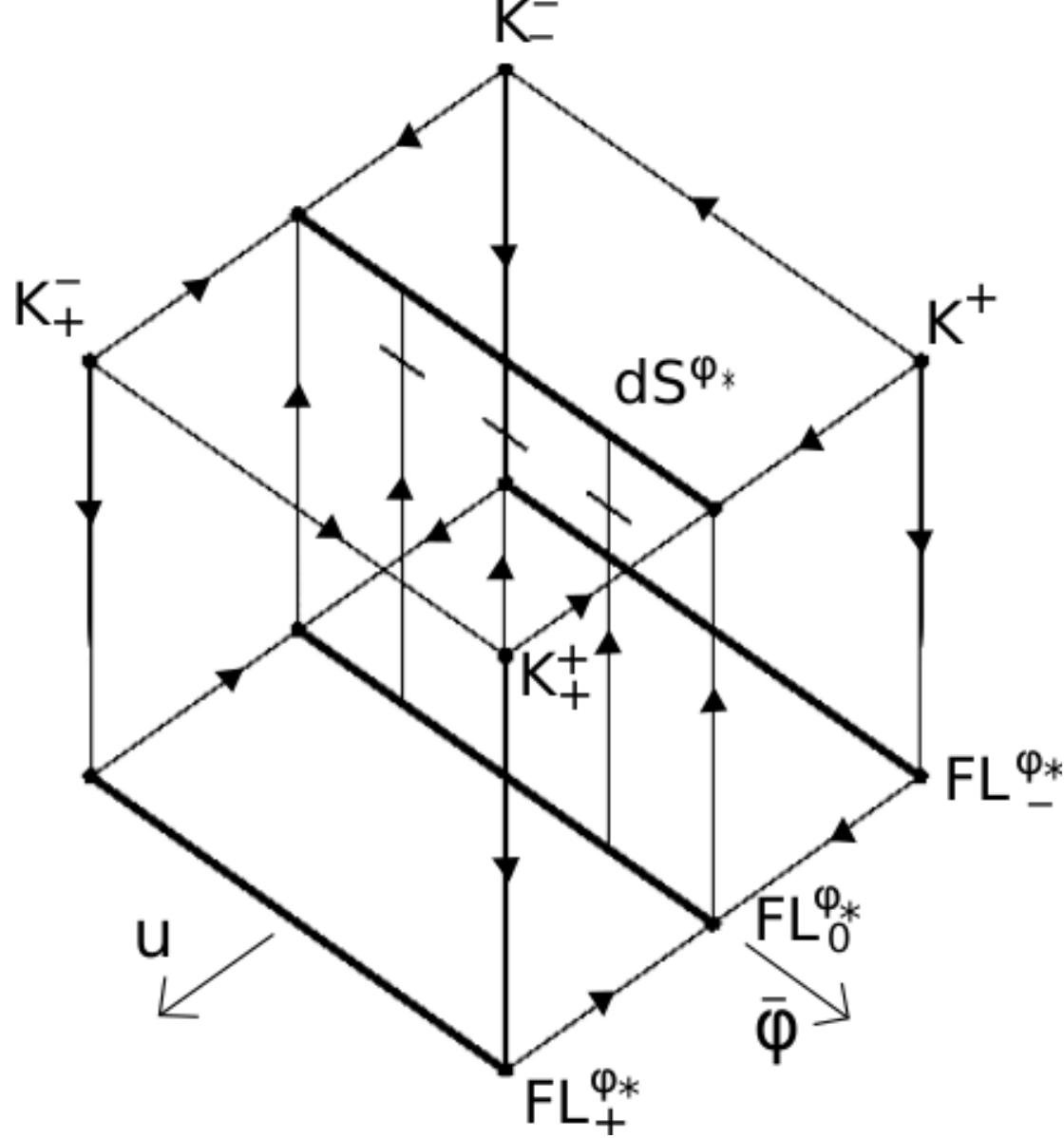}}
\hspace{0.5cm}
\subfigure[$\lambda_-=1,\,\lambda_+=0$]{\label{fig:3Dm1p0}
\includegraphics[width=0.45\textwidth]{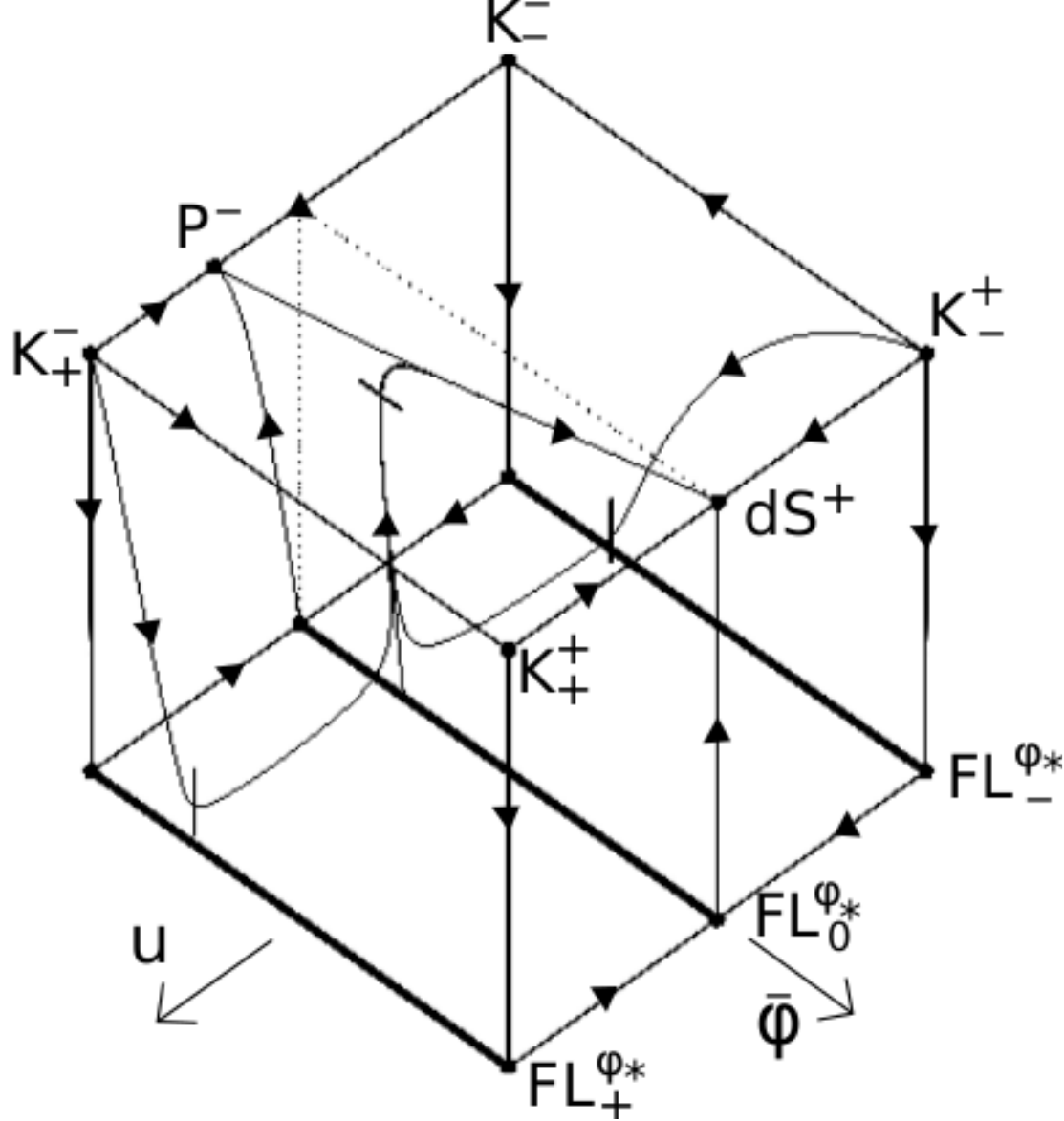}}
\hspace{0.5cm}
\subfigure[$w_\varphi(N)$ for the three orbits in (b) with $\bar{\varphi}_*=-0.5$]{\label{Figb_wDE}
\includegraphics[width=0.35\textwidth]{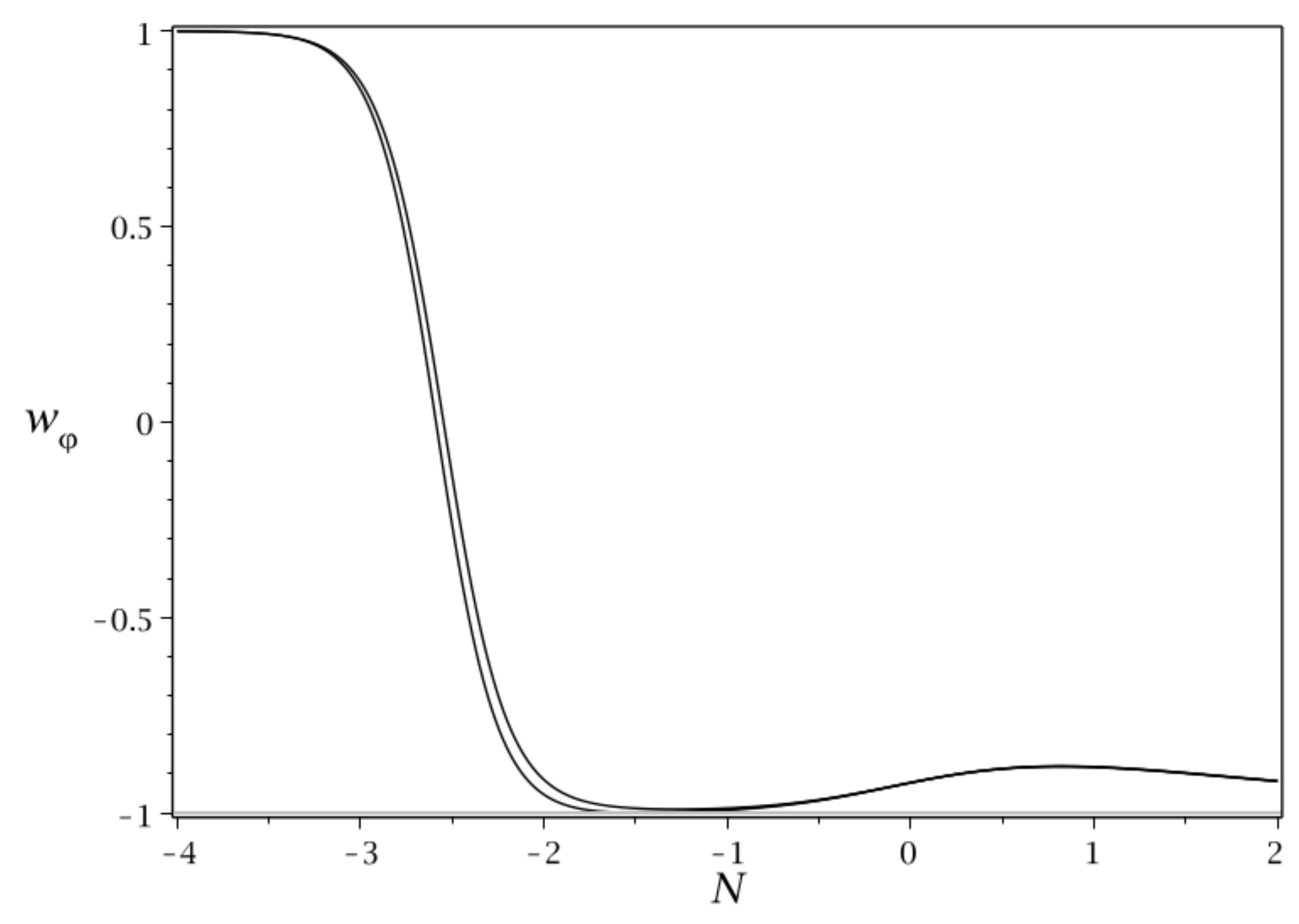}}
\hspace{0.5cm}
\subfigure[$H_{\varphi}(N)/H_\Lambda(N)$ for the three orbits in (b) with $\bar{\varphi}_*=-0.5$]{\label{Figb_E}
\includegraphics[width=0.35\textwidth]{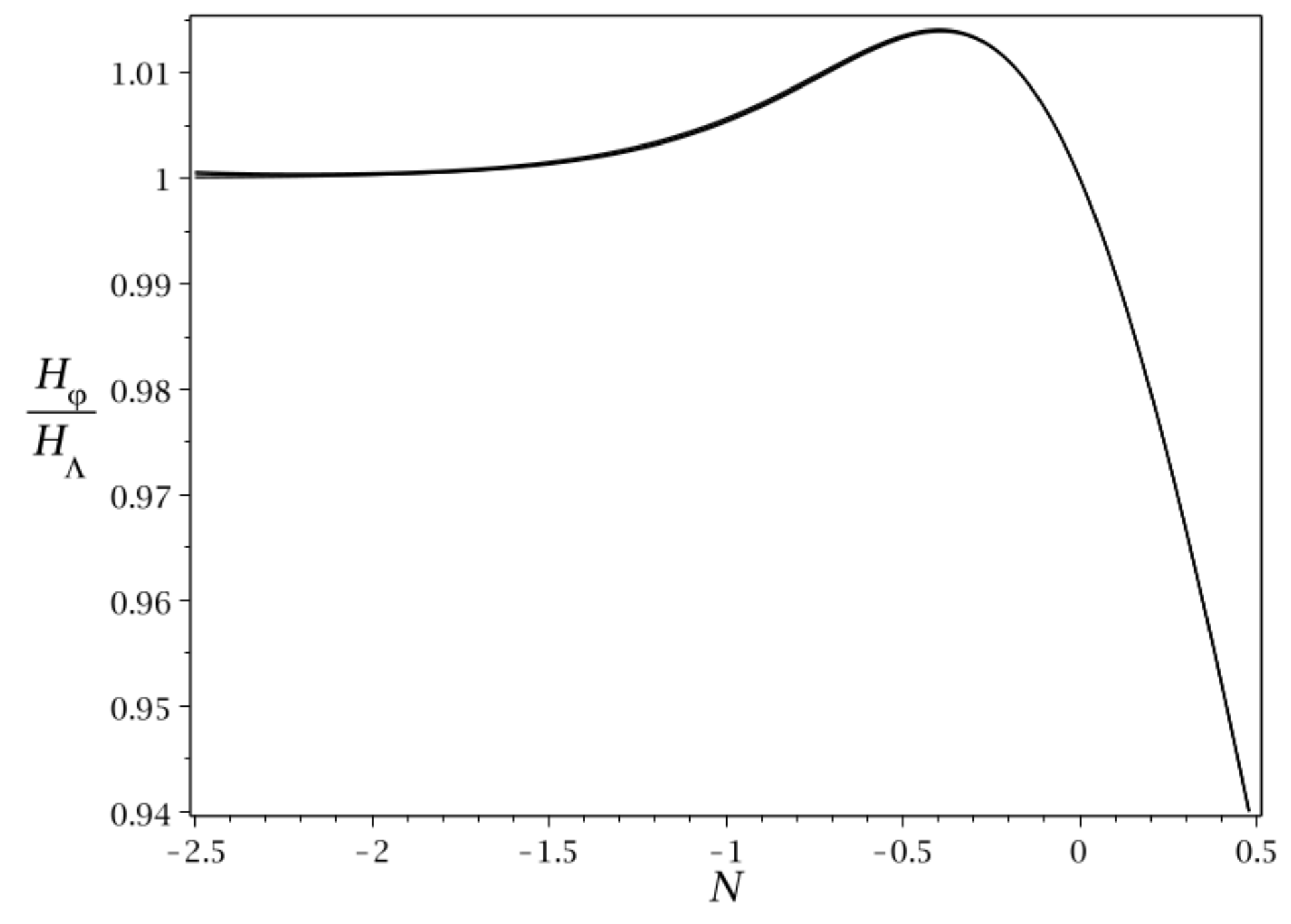}}
\vspace{-0.5cm}
\end{center}
\caption{The double-exponential
potential with $\lambda_-=1$, $\lambda_+=0$ (case 1).
Fig. (b) illustrates
the $\mathrm{FL}_0^{\varphi_*}\longrightarrow \mathrm{dS}^+$-orbit
with $\bar\varphi_*=-0.5$ and two orbits that shadow this orbit during the quintessence
epoch, where all three orbits describe thawing quintessence models.
A horizontal dash on orbits throughout denotes $N=0$ at $\Omega_\varphi=0.68$.
The matter dominated epoch is initiated slightly before the vertical dash on the orbits
which denote $N=-3$.
}\label{fig:3D00_10}
\end{figure}  
Figure~\ref{fig:3Dm1p0} depicts the $\mathrm{FL}$
unstable manifold $\mathrm{FL}_0^{\varphi_*}\rightarrow\mathrm{dS}^+$ with
$\bar\varphi_*=-0.5$, and two orbits that shadow this orbit during
its (thawing) quintessence epoch. The two
shadowing orbits, which link the past attractor to the future attractor,
are schematically described by the sequence\footnote{See the earlier discussion in
connection with~\eqref{thafree}.}
\begin{equation} \label{het.clin.seq.1.1}
\mathrm{K}_\pm^\mp\rightarrow\mathrm{FL}_\pm^{\varphi_*}\rightarrow\mathrm{FL}_0^{\varphi_*}\rightarrow\mathrm{dS}^+,\quad\text{with}\quad \bar\varphi_*=-0.5.
\end{equation}

Figure~\ref{Figb_wDE} illustrates the `bump' that appears in the graph of $w_\varphi(N)$
for all three orbits, where the maximum of the bump occurs after $N=0$. Thus all
three orbits describe \emph{thawing quintessence models}. The graphs for the shadowing orbits
show the step function behaviour of $w_\varphi(N)$ from $1$ to $-1$ that arises from the transition $\mathrm{FL}_\pm^{\varphi_*}\rightarrow\mathrm{FL}_0^{\varphi_*}$, described in section~\ref{thaw.freeze}.

Figures~\ref{fig:3Dm1p0} and~\ref{Figb_wDE} show that the two shadowing orbits have an
epoch of matter domination of relatively short duration $\Delta N \approx 2$.
This duration is determined by the minimum value of $\Omega_\varphi$ or equivalently
of $v$, which for the two shadowing orbits is given by $v_\mathrm{min} = 0.0334$.
In order to obtain a model with a more realistic value of $\Delta N \approx 8$ we have to
use shadowing orbits with $v_\mathrm{min} = 0.0004$ at $\bar{\varphi}_*=-0.5$.
This yields orbits that shadow the quintessence
$\mathrm{FL}_0^{\varphi_*}\rightarrow\mathrm{dS}^+$-orbit originating from
$\bar{\varphi}_*=-0.5$ so closely so that they are indistinguishable from it
during the quintessence epoch; moreover, before the matter dominated stage these orbits
shadow the $\mathrm{K}_\pm^\mp\rightarrow\mathrm{FL}_\pm^{\varphi_*}$ orbits with
$\bar{\varphi}_*=-0.5$ on the $u=\pm\sqrt2$ subsets, illustrated in Figure~\ref{BoxSS2},
extremely closely. This exemplifies the strong restrictions a long matter dominated epoch
imposes on initial data.
\subsection{Case 2: $\lambda_-=10$, $\lambda_+=0$\label{sec:Case2}}

\begin{figure}[ht!]    
\begin{center}
\subfigure[$\lambda_-=10,\,\lambda_+=0$]{\label{fig:3Dm10p0}
\includegraphics[width=0.45\textwidth]{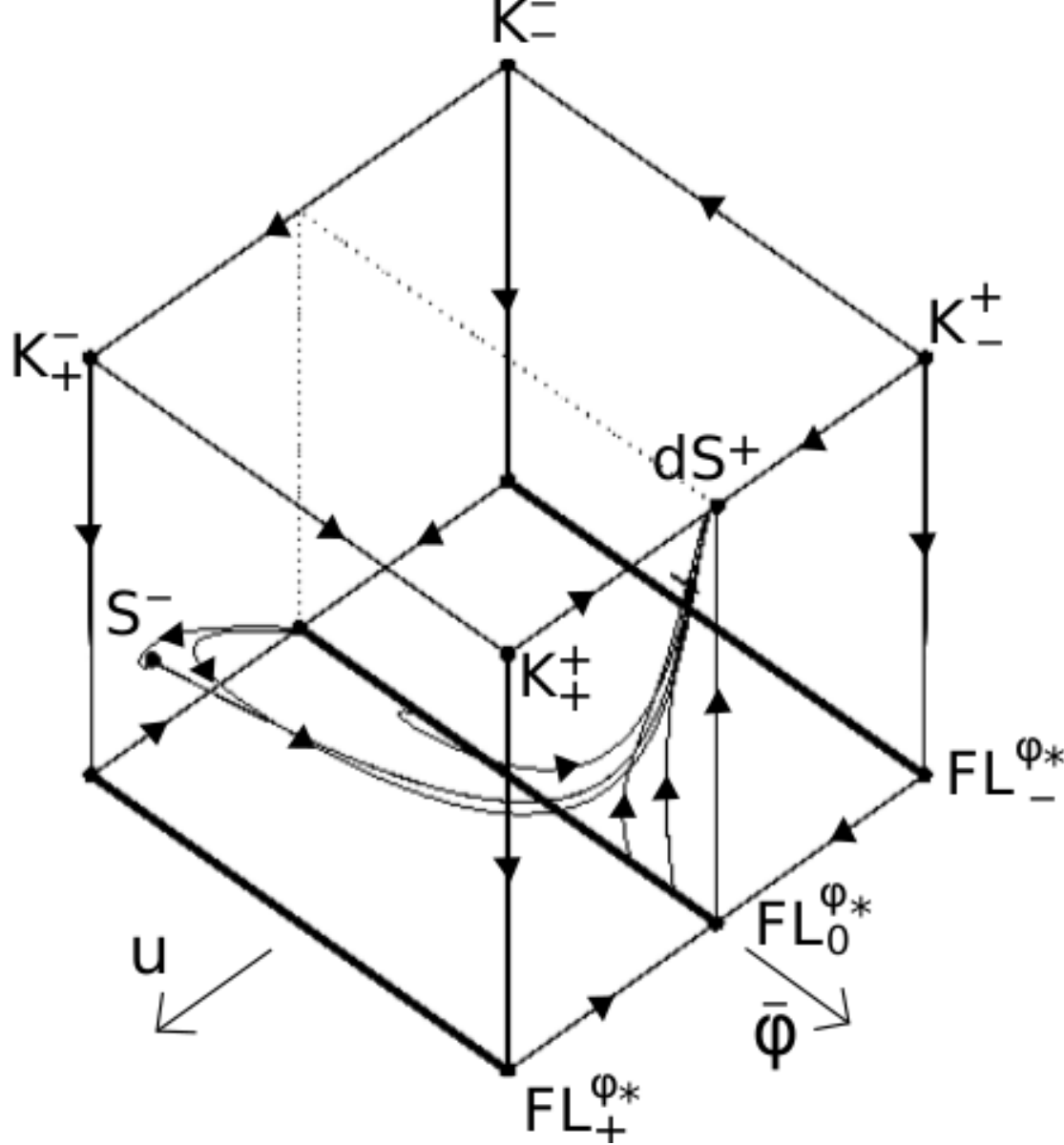}}\\
\subfigure[$w_\varphi(N)$]{\label{Figc_wDE100}
\includegraphics[width=0.35\textwidth]{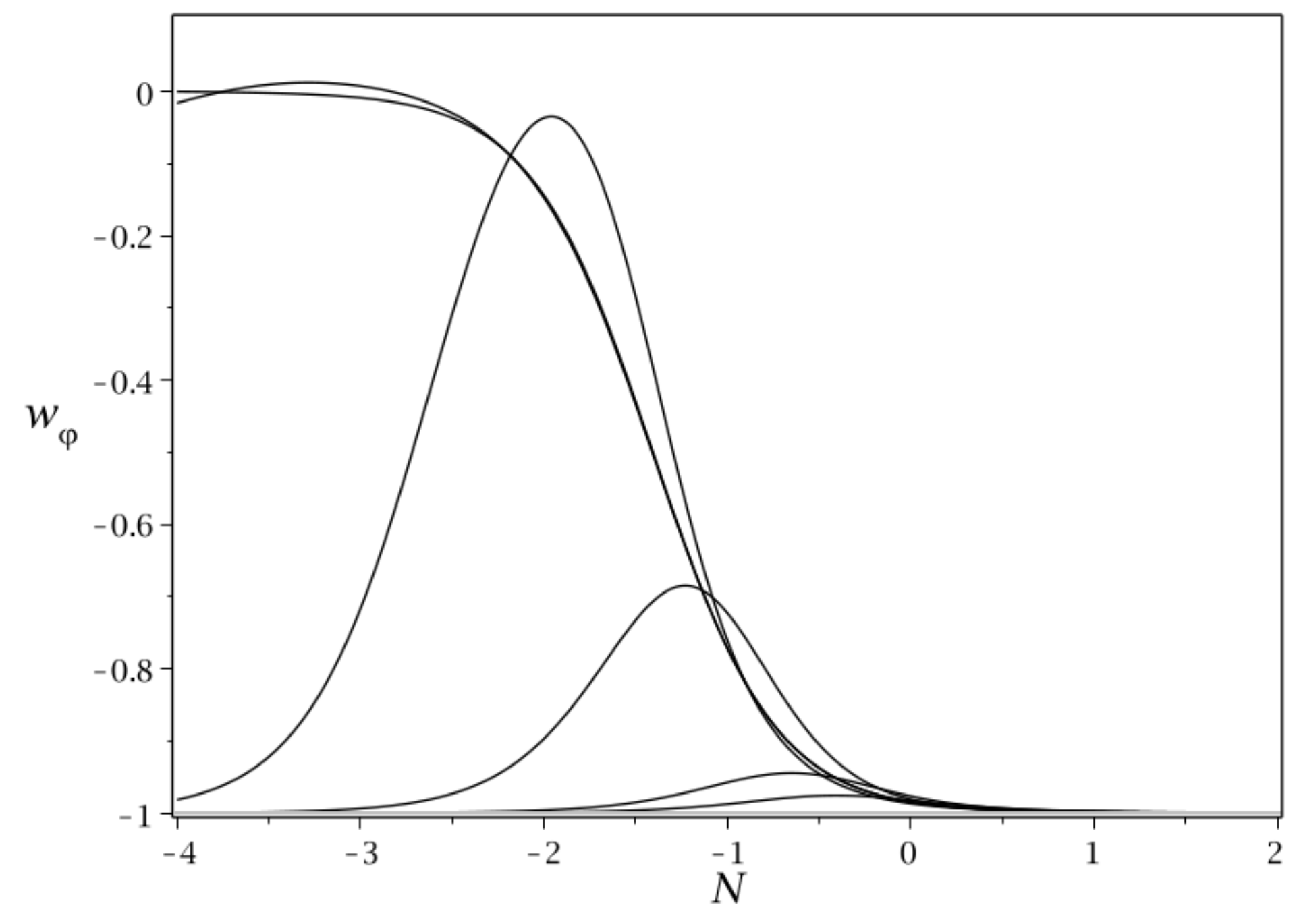}}
\hspace{0.5cm}
\subfigure[$H_\varphi(N)/H_\Lambda(N)$]{\label{Figc_E100}
\includegraphics[width=0.35\textwidth]{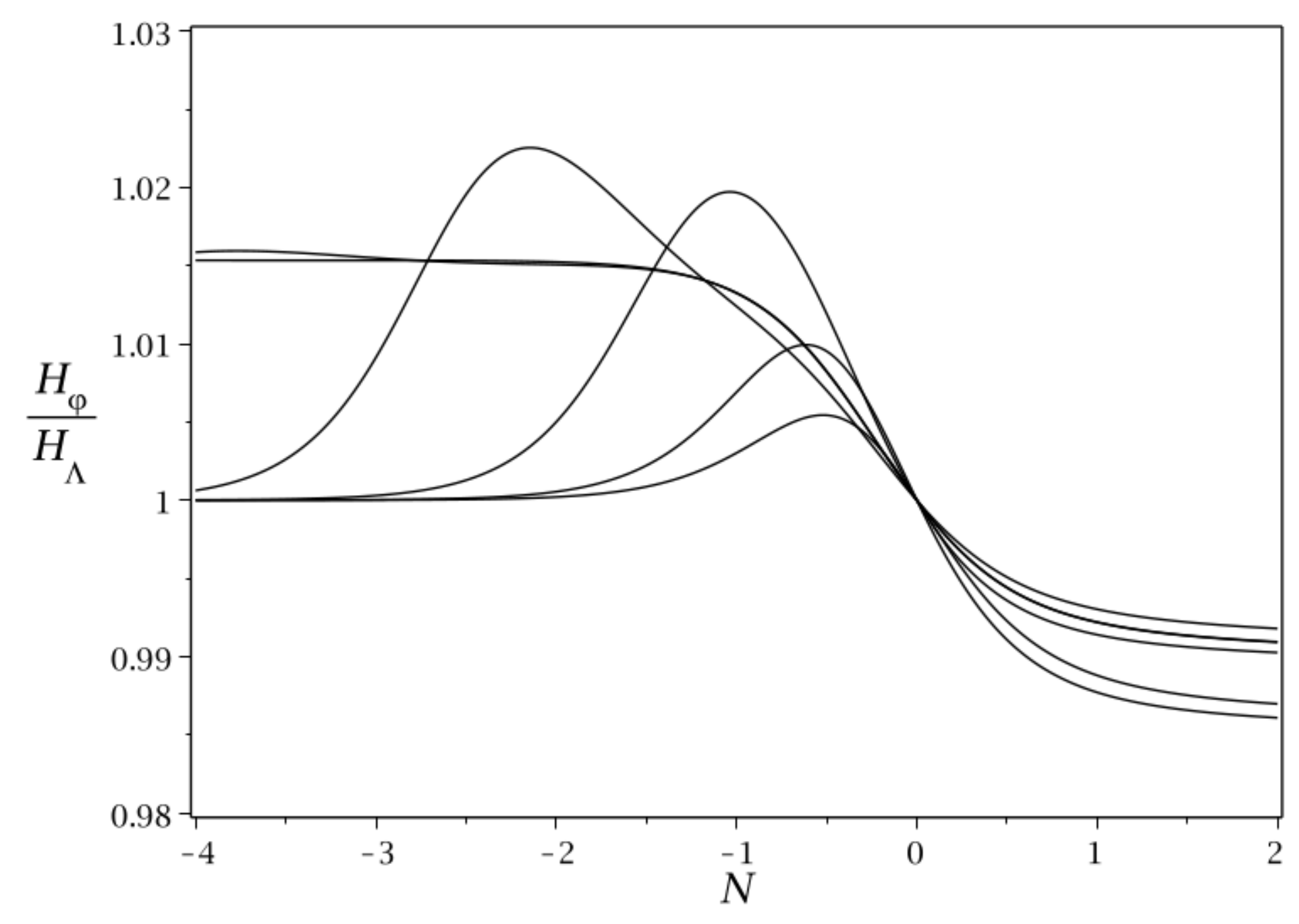}}
\vspace{-0.5cm}
\end{center}
\caption{The double-exponential
potential with $\lambda_-=10$, $\lambda_+=0$ (case 2).
Figure (a) depicts the scaling orbit from $\mathrm{S}^-$ and the
$\mathrm{FL}_0^{\varphi_*}$-orbits with $\lambda_*=1,2,7,9.9,9.99999995$,
and the corresponding values $\bar{\varphi}_*=4/5,3/5,-2/5,-0.98,-0.99999999$.
Figures (b) and (c) show the corresponding graphs for $w_\varphi(N)$ and $H_\varphi(N)/H_\Lambda(N)$.
}\label{fig:3D100}
\end{figure}

This case illustrates thawing, freezing and scaling freezing quintessence.
Figure~\ref{fig:3Dm10p0} depicts
the scaling orbit $\mathrm{S}^-\rightarrow\mathrm{dS}^+$ and five
$\mathrm{FL}_0^{\varphi_*}\rightarrow\mathrm{dS}^+$-orbits.
The $\mathrm{FL}_0^{\varphi_*}\rightarrow\mathrm{dS}^+$ orbit with
$\bar{\varphi}_* = -0.99999999$ initially shadows
the orbit $\mathrm{FL}_0^{\varphi_*}\rightarrow\mathrm{S}^-$ with $\varphi_*=-1$
very closely which brings it close enough to $\mathrm{S}^-$ so that
it has an approximate (matter dominated) scaling property when
$N\in [-4,-2.5]$ (to have longer period of scaling, e.g. $\Delta N \approx 8$,
requires orbits to be even closer to $\mathrm{S}^-$, and hence to $\bar{\varphi}=-1$);
this is followed by shadowing of the scaling orbit $\mathrm{S}^-\rightarrow\mathrm{dS}^+$
and hence is an example, as is the scaling orbit, of \emph{scaling freezing quintessence},
in accordance with section~\ref{scale.freeze}.

The remaining other $\mathrm{FL}_0^{\varphi_*}\rightarrow\mathrm{dS}^+$ orbits in
Figure~\ref{fig:3Dm10p0} illustrate the transition from thawing to freezing quintessence
as $\bar{\varphi}_*$ and thereby $\lambda_* \equiv \lambda(\bar{\varphi}_*)$ varies,
as shown by the graphs of $w_\varphi(N)$ in Figure~\ref{Figc_wDE100}. We see that
the $\mathrm{FL}_0^{\varphi_*}\rightarrow\mathrm{dS}^+$-orbits exhibit
a `bump' in $w_\varphi(N)$ with a maximum that increases and moves to more negative
$N$ as $\lambda_*$ increases: the
$\mathrm{FL}_0^{\varphi_*}\rightarrow\mathrm{dS}^+$-orbits with $\lambda_* \lesssim 7$ describe
thawing quintessence while orbits with larger values of $\lambda_*$, but not too large
since they then yield scaling freezing quintessence, describe freezing quintessence.
There are also two open sets of thawing and freezing quintessence orbits
(not shown in the figure) that shadow these
$\mathrm{FL}_0^{\varphi_*}\rightarrow\mathrm{dS}^+$-orbits during the quintessence epoch.
Before this quintessence epoch they either shadow orbits on the $u=-\sqrt{2}$ boundary and then
make the transition $\mathrm{FL}_-^{\varphi_*}\rightarrow\mathrm{FL}_0^{\varphi_*}$
or they shadow the heteroclinic sequence $\mathrm{K}_-^+\rightarrow\mathrm{K}_-^-\rightarrow\mathrm{K}_+^-$
and then orbits on the $u=\sqrt{2}$ boundary followed by the transition
$\mathrm{FL}_+^{\varphi_*}\rightarrow\mathrm{FL}_0^{\varphi_*}$. These models thus
exhibit a step like behaviour in $w_\varphi(N)$ between $+1$ and $-1$ before the
quintessence epoch, as in case 1.

Finally, Figure~\ref{Figc_E100} shows the graph $H_\varphi(N)/H_\Lambda(N)$ for the depicted
orbits in the $(\bar{\varphi},u,v)$ state space. The Hubble variable $H_\varphi(N)$ deviates
from $H_\Lambda(N)$ with less than $2\%$ during the quintessence epoch.

%

\subsection{Case 3: $\lambda_-=10$, $\lambda_+=1$\label{sec:Case3}}

\begin{figure}[ht!]  
\begin{center}
\subfigure[$\lambda_-=10,\,\lambda_+=1$]{\label{fig:3Dm10p1}
\includegraphics[width=0.45\textwidth]{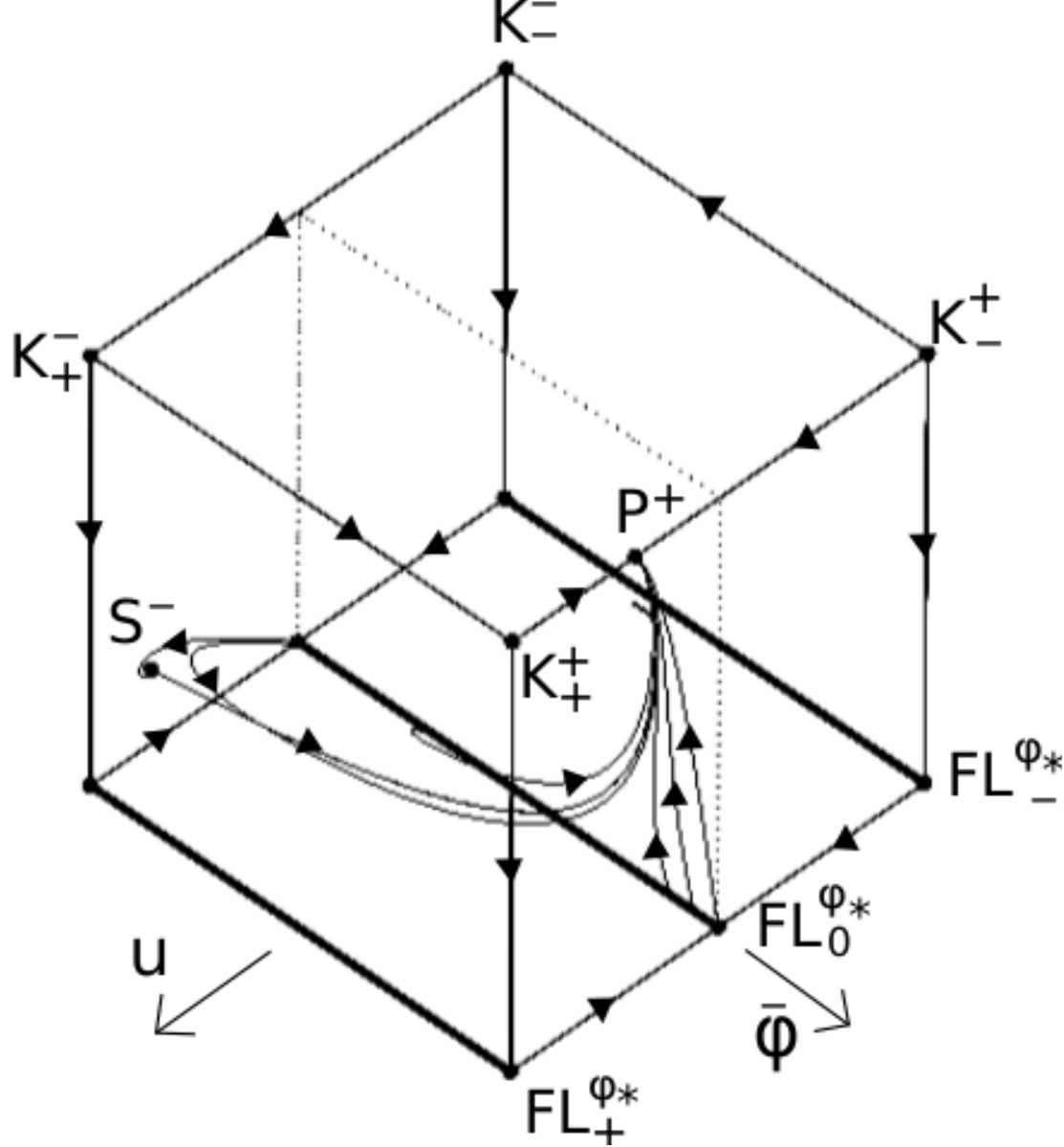}}\\
\subfigure[$w_\varphi(N)$]{\label{Figc_wDE101}
\includegraphics[width=0.35\textwidth]{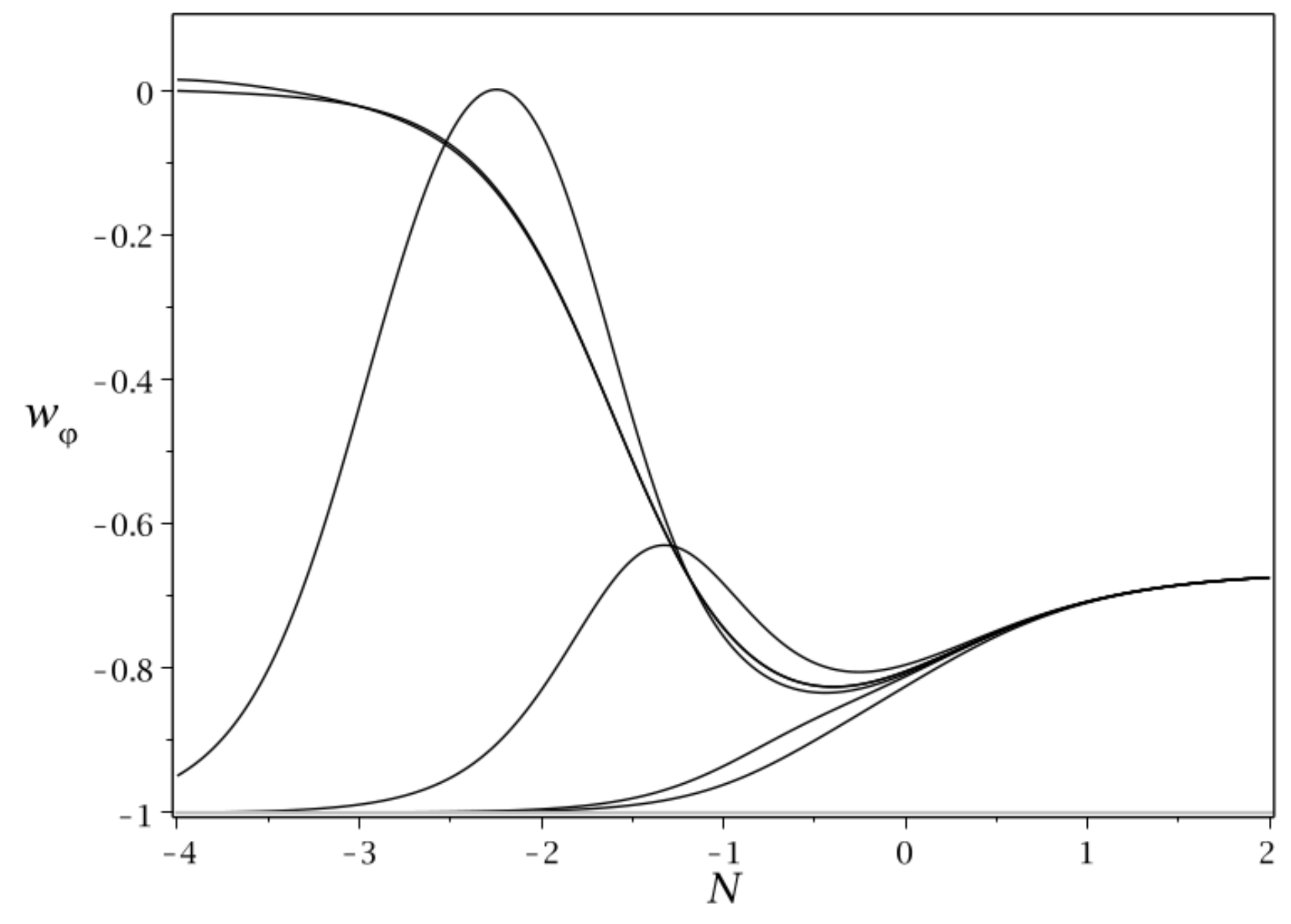}}
\hspace{0.5cm}
\subfigure[$H_\varphi(N)/H_\Lambda(N)$]{\label{Figc_E101}
\includegraphics[width=0.35\textwidth]{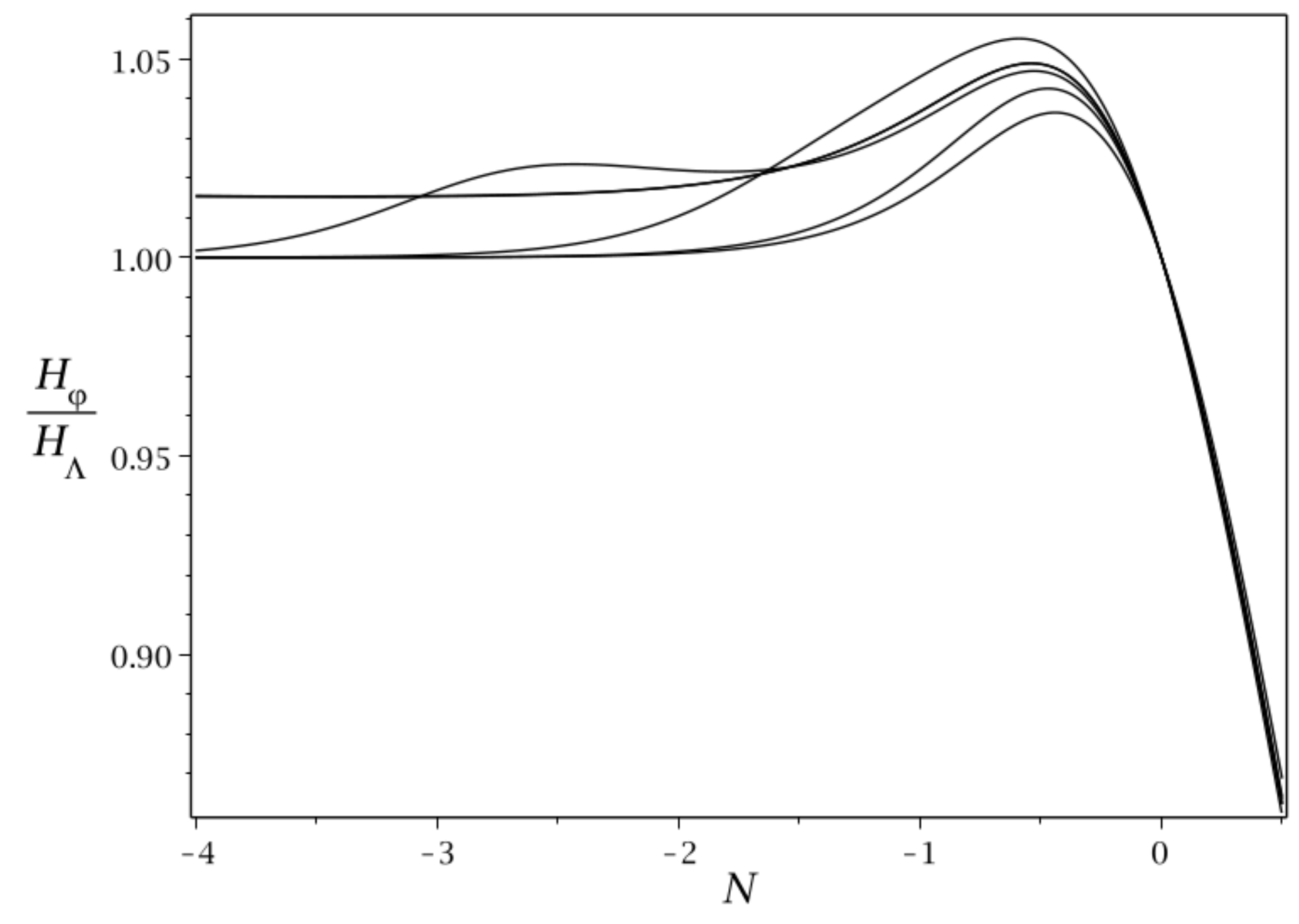}}
\vspace{-0.5cm}
\end{center}
\caption{The double-exponential potential with $\lambda_-=10, \lambda_+=1$ (case 3).
Figure (a) depicts the scaling orbit from $\mathrm{S}^-$ and the
$\mathrm{FL}_0^{\varphi_*}$-orbits with $\lambda_*=3/2,2,7,9.9,9.99999995$,
and the corresponding values $\bar{\varphi}_*=8/9, 7/9, -1/3, -0.98, -0.9999999889$.
Figures (b) and (c) show the corresponding graphs for $w_\varphi(N)$ and $H_\varphi(N)/H_\Lambda(N)$.
}\label{fig:3D101}
\end{figure}

This case illustrates the effect on the quintessence epoch of replacing $\lambda_+=0$
with a value in the range $0<\lambda_+<\sqrt{2}$, which changes the future
attractor ${\cal A}^+$ from $\mathrm{dS}^+$ to $\mathrm{P}^+$. A change from
$\lambda_+=0$ to $0<\lambda_+ \ll 1$ has negligible observational effects, but if
$\lambda_+ \approx {\cal O}(1)$ then 
$\lim_{N\rightarrow\infty}(1+w_\varphi) = \lambda_+^2/3 $ is non-negligible,
so that increasing $\lambda_+$ increasingly affects the observational 
quintessence epoch that begins at $N_\mathrm{quint} \approx -1.5$ and ends at $N=0$. 
This is illustrated by the orbits and associated graphs in Figure~\ref{fig:3D101}.
Figure~\ref{Figc_wDE101} shows that for sufficiently small $\lambda_*$ and sufficiently large
$\lambda_+$ the `bump' in $w_\varphi(N)$ is replaced by continued thawing followed by 
a plateau, while for larger $\lambda_*$ the continued freezing is replaced with 
a minimum in $w_\varphi$ where subsequent thawing is
levelled out to the asymptotically future plateau. This increases the deviation of $H_\varphi(N)$ from
$H_\Lambda(N)$ during the observational quintessence epoch so that future observations will impose increasingly
restrictive bounds on $\lambda_+$.
%
%

\subsection{Case 4: $\lambda_-=20$, $\lambda_+=-10$\label{sec:Case4}}

\begin{figure}[ht!]     
\begin{center}
\subfigure[$\lambda_-=20,\,\lambda_+=-10$]{\label{fig:3DOsc}
\includegraphics[width=0.45\textwidth]{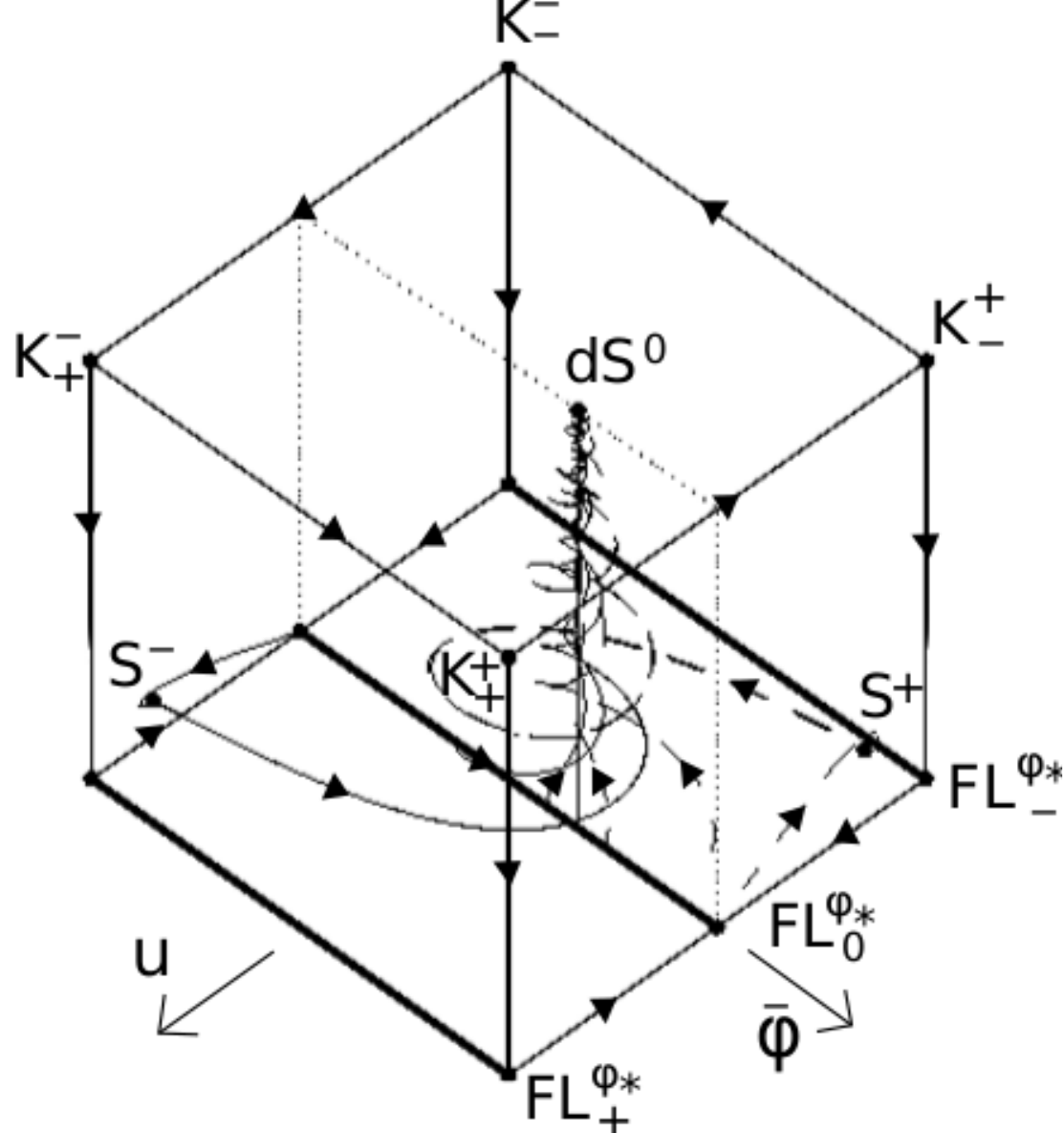}}\\
\subfigure[$w_\varphi (N)$]{\label{fig:wDE_LP}
\includegraphics[width=0.35\textwidth]{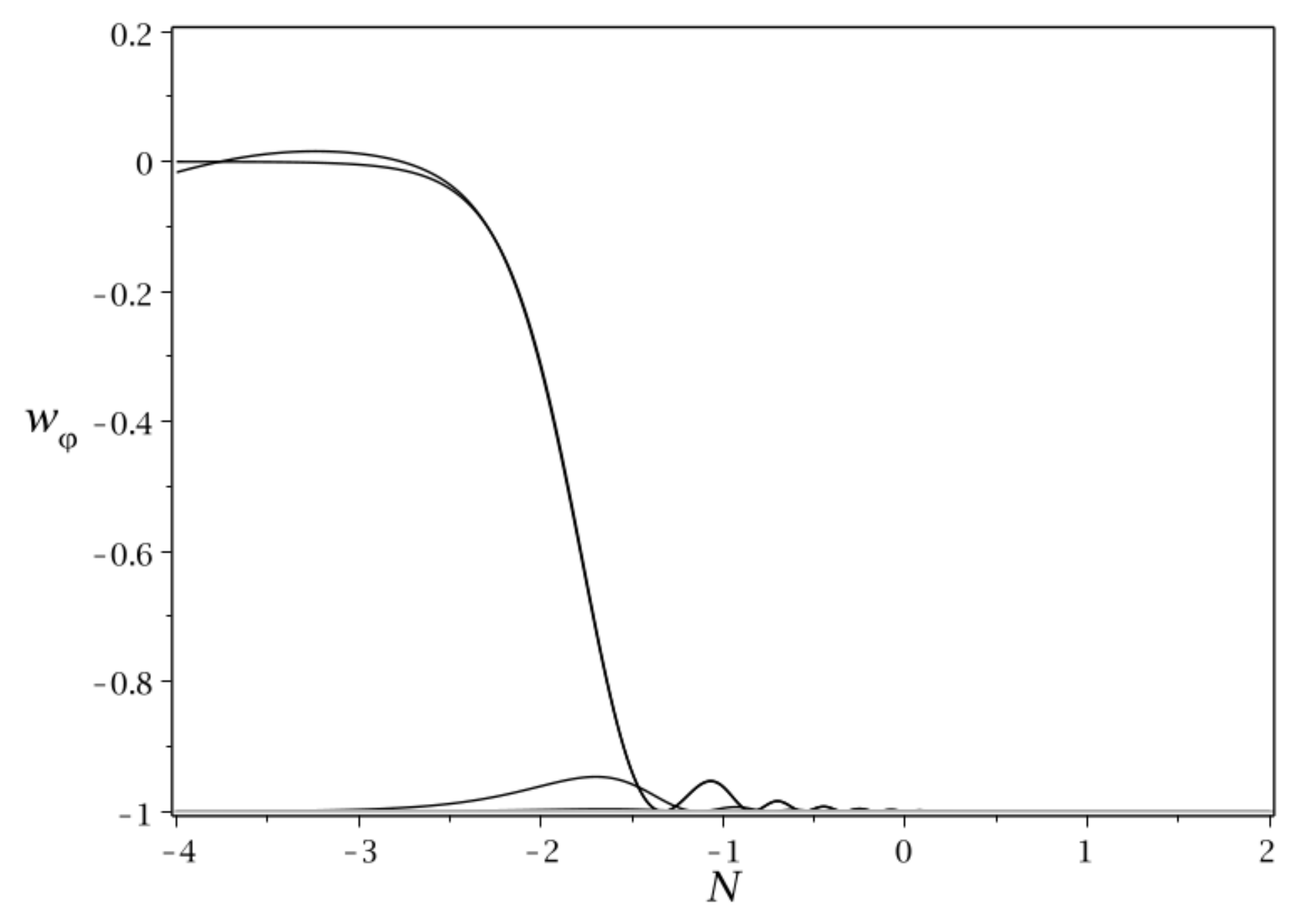}}
\hspace{0.5cm}
\subfigure[$H_{\varphi}(N)/H_\Lambda(N)$]{\label{fig:E_LP}
\includegraphics[width=0.35\textwidth]{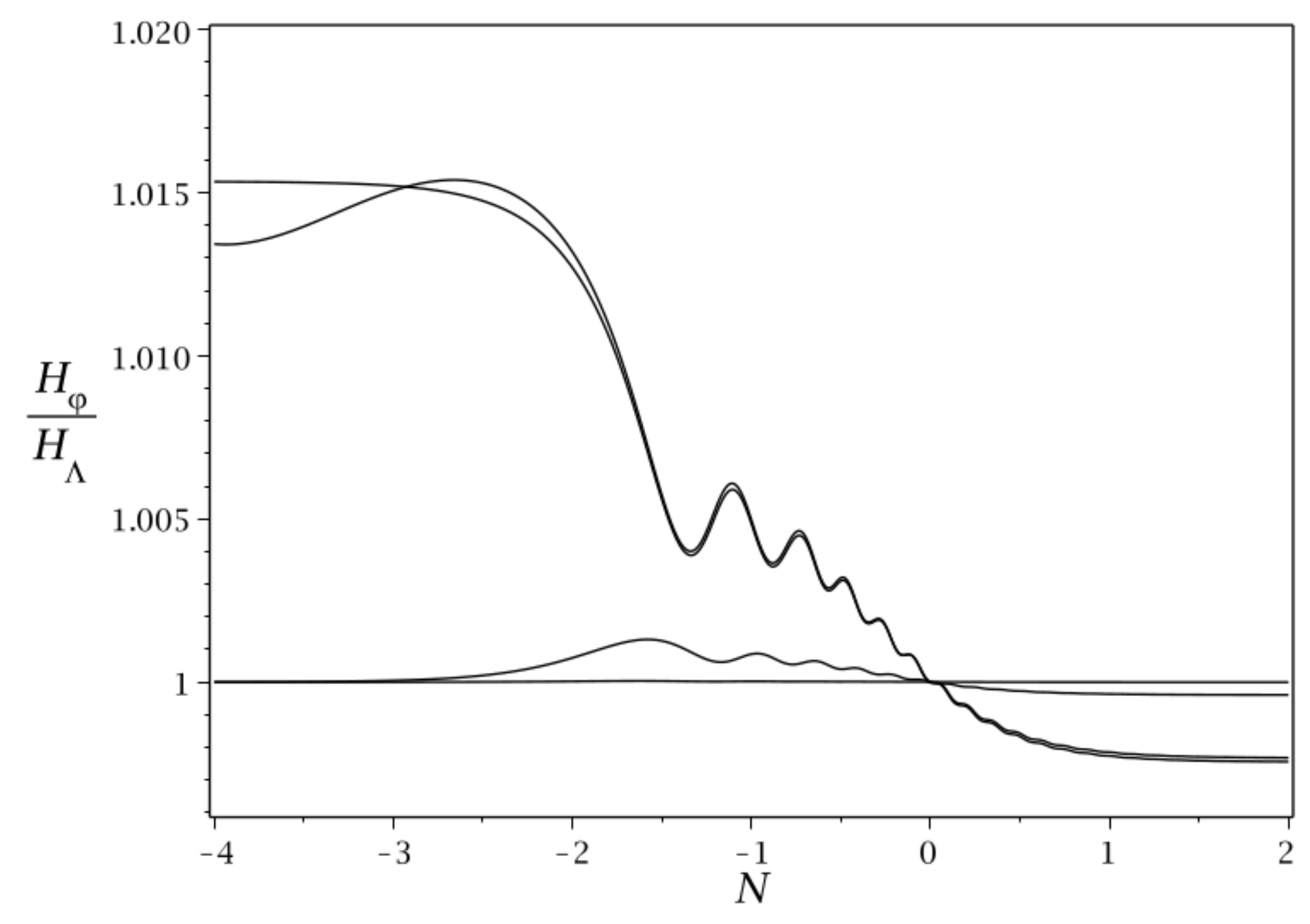}}
\vspace{-0.5cm}
\end{center}
\caption{The double-exponential potential with $\lambda_-=20, \lambda_+=-10$ (case 4).
Figure (a) shows the scaling orbits originating from $\mathrm{S}^\pm$ and various $\mathrm{FL}_0^{\varphi_*}$-orbits.
Figures (b) and (c) depict the graphs for $w_\varphi(N)$ and $H_\varphi(N)/H_\Lambda(N)$, respectively, for the
scaling orbit coming from $\mathrm{S}^-$, the $\Lambda$CDM orbit at $\bar{\varphi}_*=1/3$ with $\lambda_*=0$ and
$w_\varphi(N)=-1$, and the $\mathrm{FL}_0^{\varphi_*}$-orbits with $\lambda_*=2,8,19.99999999995$ and
$\bar{\varphi}_*=1/5,-1/5,-0.999999999997$, respectively; the $\mathrm{FL}_0^{\varphi_*}$-orbit with
$\lambda_*=2$ is indistinguishable in these graphs from the $\Lambda$CDM orbit.
}\label{fig:3D2010}
\end{figure}

This case illustrates the effect on the quintessence epoch by replacing
$\lambda_+\geq0$ with $\lambda_+ <0$. Figure~\ref{fig:3DOsc} shows the
scaling orbits originating from $\mathrm{S}^-$ and $\mathrm{S}^+$ and some
$\mathrm{FL}_0^{\varphi_*}\rightarrow\mathrm{dS}^0$-orbits. The scaling orbits
are asymptotic to the fixed point $\mathrm{dS}^{0}$ and spiral around the
straight line $\Lambda$CDM orbit as they approach $\mathrm{dS}^{0}$. The
scaling orbits describe models with a scaling phase in part of the matter
dominated epoch followed by oscillations of $w_\varphi$. For this reason
we referred to this type of models as \emph{scaling oscillatory quintessence}
models, in section~\ref{sec:scale.osc}. The orbits that originate from
$\mathrm{FL}_0^{\varphi_*}$ with $\bar{\varphi}_*$ extremely close to $-1$
(respectively $+1$) shadow the scaling orbit from $\mathrm{S}^-$
(respectively $\mathrm{S}^+$) and hence also describe scaling oscillatory
quintessence (as does other, non-illustrated, open sets of orbits that shadow
orbits on the $\bar{\varphi}=\pm 1$ boundaries and come extremely close
to $\mathrm{S}^\pm$). In contrast, the orbits
$\mathrm{FL}_0^{\varphi_*}\rightarrow\mathrm{dS}^0$-orbits in~\ref{fig:3DOsc}
with $\bar{\varphi}_*$ that are not extremely close to $\pm1$
do not have a scaling phase and are examples of models that we referred to as
\emph{oscillatory quintessence} models in section~\ref{sec:scale.osc}.

Figures~\ref{fig:wDE_LP} and~\ref{fig:E_LP} show the graphs of $w_\varphi$ and
$H_\varphi(N)/H_\Lambda(N)$ for values of $\bar\varphi_*>\bar\varphi_0$.
Figure~\ref{fig:wDE_LP} shows the familiar steep drop in $w_\varphi$ from
the plateaux at $w_\varphi \approx 0$ that characterizes scaling oscillatory
quintessence.\footnote{Increasing $\lambda_-$ moves the steep drop in $w_\varphi$
for the scaling (freezing) orbit (scaling (oscillatory) orbit) in Figure~\ref{Figc_wDE100}
(Figure~\ref{fig:wDE_LP}) to increasingly negative $N$, in agreement with Figures 5
and 9 in Bag {\it et al}. (2018)~\cite{bagetal18}.} Figure~\ref{fig:E_LP} shows
that $H_\varphi(N)$ deviates from $H_\Lambda(N)$ very little during the quintessence epoch,
for both scaling oscillatory quintessence and oscillatory quintessence.
When compared with Figure~\ref{Figc_E100}\footnote{For example, the
graph in Figure~\ref{fig:E_LP} with $\lambda_*=8$ has $H_\varphi(N)/H_\Lambda(N)\lesssim 1.002$, while
the graph in Figure~\ref{Figc_E100} with $\lambda_*=7$ has $H_\varphi(N)/H_\Lambda(N)\lesssim1.02$.}
this suggests that potentials with a positive minimum and $\lambda_-\gg1$ yield quintessence evolution
that is closer to $\Lambda$CDM evolution than monotonic potentials for which $\lambda_-\gg1$.

\section{Concluding remarks \label{concl.remarks}}

%
%

In this paper we have analyzed different types of quintessence that arise for scalar
field potentials for which $\lambda(\varphi)$ is bounded with limits
$\lambda_{\pm}= \lim_{\varphi\rightarrow \pm\infty}\lambda(\varphi)$, using a new regular
dynamical system on a three-dimensional bounded `box state space' $(\bar{\varphi},u,v)$,
illustrated with specific double exponential potential examples. The box state space
made it possible to systematically explore the entire solution space of models with
bounded $\lambda(\varphi)$. This new formulation also highlighted that quintessence 
dynamics is described by the one-dimensional unstable manifold orbits of the
fixed points $\mathrm{FL}_0^{\varphi_*}$ and $\mathrm{S}^{\pm}$. These orbits
describe the dynamics during the quintessence epoch, when $\Omega_\varphi$ is
increasing and beginning to influence the cosmological expansion. In order to 
model the matter-dominated epoch which precedes this epoch one has to consider
an open set of orbits that
originate at the past attractor and shadow the unstable manifolds during the
quintessence epoch. The requirement of a sufficiently long matter dominated epoch
(of order $8 \,e$-folds) requires orbits to come very close to one or
more of the fixed points $\mathrm{S}^\pm$, $\mathrm{FL}_0^{\varphi_*}$ in the matter
dominated part of the state space. This severely restricts viable quintessence
initial data but has the advantage that the open set of orbits subsequently
shadow the unstable manifold orbits very closely, which thereby describe quintessence.


Analysis of the unstable manifold orbits revealed a rather wide range of dynamical
possibilities during the quintessence epoch. This resulted in an extension of the quintessence
classification of Tsujikawa (2013)~\cite{tsu13} for models with bounded $\lambda(\varphi)$
from thawing quintessence and scaling freezing quintessence to also include freezing
quintessence, scaling oscillatory quintessence and oscillatory quintessence, where the
last two types occur for potentials with a sufficiently steep positive minimum ({\it i.e.}
potentials with $\lambda_{,\varphi}(\bar{\varphi}_0)<-3/4$).

More precisely, thawing quintessence evolution is described by rolling down a monotonic
potential according to $\mathrm{FL}_0^{\varphi_*} \rightarrow {\cal A}^+$ when
$\lambda_* = O(1)$, while freezing quintessence arises when $\lambda_* > O(1)$ is large,
but avoiding a large $\lambda_-$ (recall the general $w_\varphi$ `bump' discussion
in section~\ref{sec:monotonic} exemplified and illustrated in sections~\ref{sec:Case1},
\ref{sec:Case2} and~\ref{sec:Case3}); scaling freezing quintessence evolution
is described by rolling down a monotonic potential according to $\mathrm{S}^- \rightarrow {\cal A}^+$,
see section~\ref{sec:monotonic} and sections~\ref{sec:Case2} and~\ref{sec:Case3} for illustrations
(in all these cases ${\cal A}^+=\mathrm{dS}^+$ when $\lambda_+=0$, ${\cal A}^+=\mathrm{P}^+$ when
$0<\lambda_+<\sqrt{2}$). Scaling oscillatory quintessence and oscillatory quintessence evolution
arise for potentials with a minimum where $\lambda_{,\varphi}<-3/4$, where scaling oscillatory
quintessence evolution is described by the scaling orbits $\mathrm{S}^\pm\rightarrow\mathrm{dS}^0$
and orbits that shadow them extremely closely, thereby allowing scaling in part of the matter
dominated epoch, while oscillatory quintessence evolution corresponds to orbits
$\mathrm{FL}_0^{\varphi_*}\rightarrow\mathrm{dS}^0$ and an open set of orbits shadowing
these orbits, where $\bar{\varphi}_*$ is not too close to $\bar{\varphi}=\pm 1$; see
section~\ref{sec:scale.osc} for the general discussion and section~\ref{sec:Case4} for
examples and illustrations.

These various types of quintessence evolution were illustrated in section~\ref{sec:doubleexp}, but
there are also some illustrative results in the literature:
\begin{itemize}
\item[i)] Thawing quintessence: See
Figures~\ref{fig:3D00_10}, \ref{fig:3D100} and~\ref{fig:3D101} and also
Akrami {\it et al.} (2020)~\cite{akretal20}, Figure 1, top (lower) panel, with
$\lambda_+=0$ ($\lambda_+>0$).\footnote{There are also examples in the literature where
thawing quintessence corresponds to slowly rolling down a \emph{local} monotonic \emph{part} of
a potential, {\it e.g.} the hilltop potential $V=V_0(1+\sech(\alpha\varphi/f)$ (which can be
treated globally with the present formulation since $\lambda(\varphi)$ is bounded, although
we, for brevity, omit to do so), see Figure 2 in Yang {\it et al.} 2019~\cite{yanetal19}. Another
example is the PNGB potential $V=V_0(1+\cos(\varphi/f)$, which describes thawing quintessence
as long as the scalar field is slowly rolling down a potential slope (see, {\it e.g.},
Tsujikawa (2013)~\cite{tsu13} section 3.3). In this case, however, the time period for thawing
is limited since the slow roll is interrupted by oscillations at the potential minimum. Due
to that $\lambda\rightarrow \infty$ at the minimum, this potential is not globally covered by
the present formulation, but the oscillatory evolution can be described by using a
modification of the methods given by Alho {\it et al}. (2015)~\cite{alhetal15}.}
\item[ii)] Freezing quintessence: See Figures~\ref{fig:3D100} and~\ref{fig:3D101}.
We are not aware of papers that discuss this case.
\item[iii)] Scaling freezing quintessence: See Figures~\ref{fig:3D100} and~\ref{fig:3D101};
Barreiro {\it et al.} (2000)~\cite{baretal00}, Figure 3;
Bassett {\it et al.} (2008)~\cite{basetal08}, Figure 2; Chiba {\it et al.} (2013)~\cite{chietal13},
Figures 2 and 3; Bag {\it et al.} (2018)~\cite{bagetal18}, Figure 9.
\item[iv)] Scaling oscillatory quintessence: See Figure~\ref{fig:3D2010};
Barreiro {\it et al.} (2000)~\cite{baretal00}, Figure 3;\footnote{This paper shows
$w_\varphi(N)$ on the same axes for two cases of the double exponential potential,
$(\lambda_+,\lambda_-)=(20,0.5)$, scaling freezing quintessence; $(\lambda_+,\lambda_-)=(20,-20)$,
scaling oscillatory quintessence.}
Bassett {\it et al.} (2008)~\cite{basetal08}, figure 2;
Bag {\it et al.} (2018)~\cite{bagetal18}, Figure 5.
\item[v)] Oscillatory quintessence: See Figure~\ref{fig:3D2010};
Yang {\it et al.} 2019~\cite{yanetal19}, Figure 1.\footnote{The graph of $w_\varphi(N)$ shows a drop
from $1$ to $-1$ (corresponding to initial data close to $\mathrm{FL}_-^{\varphi_*}$ followed
by shadowing $\mathrm{FL}_-^{\varphi_*}\rightarrow\mathrm{FL}_0^{\varphi_*}$) and subsequent
oscillations with an initial large `bump' with an amplitude $\approx1$. There are no radiation
or dust plateaus at $1/3$ and $0$, respectively for $w_\varphi$ and the model thereby have no
radiation or dust scaling epochs.}
\end{itemize}

In the present work we have for simplicity restricted matter to be dust, but in future work we will
include radiation in an extended state space description, which have the present state space as an
invariant dust boundary, and an analogous invariant radiation boundary where the dust content is
zero. For large $\lambda_-$ (and $-\lambda_+$ in the potential minimum case) this leads to two
scaling plateaus for scaling freezing quintessence (and scaling oscillating quintessence):
one for radiation at $w_\varphi(N)= 1/3$, corresponding to orbits originating from, or coming
very closely to, a scaling fixed point on the radiation boundary,
and one plateaux at $w_\varphi(N)= 0$, associated with that the radiation scaling orbit(s) come extremely
close to the scaling fixed point at the dust boundary after the radiation-dust transition. This gives
a dynamical systems description of these two plateaus in the graphs $w_\varphi(N)$ for scaling freezing
and scaling oscillatory quintessence for the radiation and dust case in the literature, see
Barreiro {\it et al.} (2000)~\cite{baretal00}, Figure 3, and Bag {\it et al.} (2018)~\cite{bagetal18},
Figures 5 and 9.

As mentioned in the introduction,
Tsujikawa (2013)~\cite{tsu13}, following Steinhardt {\it et al} (1999)~\cite{steetal99},
defined tracking freezing quintessence for potentials which
satisfy $\lim_{\varphi \rightarrow 0}\lambda=\infty$, {\it e.g.} the inverse power law potential.
\footnote{Steinhardt {\it et al}. (1999)~\cite{steetal99} introduced the term `tracking solutions' 
to describe this form of quintessence (a wide range of initial conditions rapidly converge
to a common evolutionary track). Tsujikawa added the qualifier `freezing' since
$w_\varphi'<0$. For brevity we will use the name tracking quintessence.}
For such potentials the new dynamical system~\eqref{Dynsys.uv} is not regular, since $\lambda(\varphi)$
appears on the right hand side of the equations. We have recently found an alternate set of
$\bar{\varphi},u,v$ variables that overcomes this difficulty, {\it i.e.} the resulting
dynamical system is regular, but at the expense of a new variable $v$ that is unbounded.
This alternate system has enabled us to describe tracking quintessence from a state space
perspective in terms of the unstable manifold of a matter dominated `tracking'
fixed point, which we refer to as the `tracking orbit',
and an open set of nearby orbits that track, {\it i.e.}
shadow, the tracking orbit.\footnote{There is an analogy with the scaling orbit and the
representation of scaling freezing quintessence in the present paper.}
Although the state space is unbounded, the tracking orbit
and the open set of shadowing orbits during matter domination and quintessence evolution,
are confined to a bounded region of the state space. Our treatment of tracking
quintessence will be given in a subsequent paper.

In yet another paper we will use the new `tracking dynamical system' \emph{and} the
`tent state space' $(\bar{\varphi}, \Sigma_\varphi, \Omega_\mathrm{m})$ formulation
to derive new, simple, and accurate approximation formulas
for key quantities such as $w_\varphi(N)$ and $H_\varphi(N)$ in a systematic and
unified manner for the various types of quintessence, thereby complementing earlier
work in the literature.\footnote{See, for example~\cite{tsu13}, section 3 for a review.}


\section*{Acknowledgments}
AA is supported by FCT/Portugal through CAMGSD, IST-ID, Projects No. UIDB/04459/2020 and No. UIDP/04459/2020. CU would like to thank the CAMGSD, Instituto Superior T\'ecnico in Lisbon, Portugal, for kind hospitality.

\bibliographystyle{unsrt}
\bibliography{../Bibtex/cos_pert_papers}

\end{document}